\documentclass[11pt]{article}
\usepackage{cite}
\usepackage{amsmath,amssymb,amsfonts}
\usepackage{algorithmic}
\usepackage{graphicx}
\usepackage{textcomp}
\usepackage[percent]{overpic}

\newtheorem{lemma}{Lemma}

\newtheorem{prop}{Proposition}

\newtheorem{defi}{Definition}
 \newtheorem{proper}{Property}

\begin{document}
\begin{center}
\Large{Non-Asymptotic State and Disturbance Estimation for a Class of  Triangular Nonlinear Systems using Modulating Functions}\\

Yasmine Marani$^{1}$, Ibrahima N'Doye$^{1}$, and Taous-Meriem Laleg-Kirati$^{2,1}$ \\
\end{center}

{$^{1}$Computer, Electrical and Mathematical Science and Engineering Division (CEMSE), King Abdullah University of Science and Technology (KAUST), Thuwal 23955-6900, Saudi Arabia (e-mail: yasmine.marani@kaust.edu.sa; ibrahima.ndoye@kaust.edu.sa; taousmeriem.laleg@kaust.edu.sa)}

{$^{2}$National Institute for Research in Digital Science and Technology, Paris-Saclay, France.}



\begin{abstract}
Dynamical models  are often corrupted by model uncertainties, external disturbances, and measurement noise. These factors affect the performance of model-based observers and as a result, affect the closed-loop performance. Therefore, it is critical to develop robust model-based estimators that reconstruct both the states and the model disturbances while mitigating the effect of measurement noise in order to ensure good system monitoring and closed-loop performance when designing controllers.  
In this article, a robust step by step non-asymptotic observer for triangular nonlinear systems for the joint estimation of the state and the disturbance is developed. The proposed approach provides a sequential estimation of the states and the disturbance in finite time using smooth modulating functions. The robustness of the proposed observer is both in the sense of model disturbances and measurement noise. In fact, the structure of triangular systems combined with the modulating function-based method allows the estimation of the states independently of model disturbances and the integral operator involved in the modulating function-based method mitigates the noise. Additionally, the modulating function method shifts the derivative from the noisy output to the smooth modulating function which strengthens its robustness properties. The applicability of the proposed modulating function-based estimator is illustrated in numerical simulations and compared to a second-order sliding mode super twisting observer under different measurement noise levels.

\end{abstract}

\paragraph{\textbf{Key words} }
Modulating functions, non-asymptotic estimation, nonlinear systems, disturbance estimation.

\section{Introduction}
\label{sec:introduction}
\noindent

\noindent Physical systems can be described by mathematical models that are often subject to disturbances caused by modeling uncertainties, faults, or mutations in the system's behavior \cite{Koen2005}. In bioreactors, the mathematical model is subject to model uncertainty in the form of a partially known nonlinear function of the state called the reaction rate \cite{Kleinstreuer2006}. Another example is the binary distillation column systems in which the dynamics of the feed compositions is unknown \cite{Hammouri2002}. The presence of unknown model disturbances and measurement noise affect the state estimation and make the model based observer design task challenging. As a result, the closed-loop performance of the system is affected. That being the case, a robust estimation algorithm that jointly estimates the states and the disturbance is required to ensure good closed-loop performance.\\

\noindent
The simultaneous estimation of the states and the disturbance for nonlinear systems affected by model disturbances has been addressed in a number of papers. For instance, an Extended Kalman Filter with Unknown Input (EKF-UI) is developed in \cite{Simone2006} and in \cite{Pertew2005} an adaptive observer for a nonlinear system with Lipschitz non-linearities is proposed. Both these observers impose strong assumptions on the disturbance (unknown input), where it is assumed to be piecewise constant. In addition the EKF-UI does not provide global convergence guarantees. Other works take another direction when dealing with such systems by using learning-based methods.  Neural networks are used to approximate the model disturbance  \cite{KIM19971539}, \cite{Vargas2000} \cite{Ren2018}. However these learning based observers require training of the neural network which prevents their applicability in online estimation and control framework. In addition, these neuro-based observers suffer from all the drawbacks of learning based approaches such as generalization and sensitivity to noise. However, most of the above-mentioned methods provide an asymptotic estimation of the state and often do not reconstruct the disturbance term. In most cases, non-asymptotic or finite time estimation is important, especially for nonlinear systems as it allows to satisfy the separation principle.\\

\noindent
A number of finite time observers for nonlinear systems subject to model disturbance have been reported in the literature. The most common one is the sliding mode observer \cite{Yi2001}. For instance, a sliding mode observer for a class of uncertain triangular systems was developed in \cite{Ahmed1999}. In \cite{Fridman2005}, a second-order sliding mode super twisting observer was proposed for mechanical systems which was generalized to a larger class of systems in \cite{SALGADO2011}. Despite providing finite time convergence properties of the states in presence of uncertainties and disturbances, these observers do not allow the estimation  of the disturbance term. \\

\noindent
Recently, a new approach for non-asymptotic state estimation called Modulating Function Based Method (MFBM) was proposed (see, for instance \cite{Jouffroy2015,korder2022,Yasmine2023,ANL:23,GHAFFOUR2023}). MFBM has desired properties of common interest that include non-asymptotic convergence and robustness features. Fortunately, it has the capability to transform the ordinary differential equations into a set of algebraic equations by means of an integral operator. Solving this algebraic system prevents solving the direct problem which allows reconstructing the states in finite time without requiring initial conditions.  
MFBM was originally used in the fifties for parameter identification problems of Ordinary Differential Equations (ODEs) \cite{shinbrot1957}. The method was then extended to source and parameter estimation of one-dimensional PDEs \cite{Sharefa2015}, \cite{Sharefa2017}, \cite{Fischer2018} and fractional order differential equations  \cite{Liu2013, Abeer2015, Dayan2023, Liu2021}. The MFBM was used for state estimation for the first time in \cite{Jouffroy2015} for linear ODEs.  Later, it was extended to linear PDEs \cite{GHAFFOUR2020} and more recently to nonlinear PDEs \cite{GHAFFOUR2023}. The modulating function-based approach was also used to estimate the pseudo-state of a fractional linear differential equation \cite{Liu2017}. Nevertheless, the design of modulating function-based estimators for nonlinear ODEs is still an active research area and few works exist in the literature. For instance, in \cite{korder2022} a linearization-based modulating function observer for a class of nonlinear systems is proposed.  In \cite{DBA:19}, the authors propose a coordinate transformation based on MFBM that transforms the original system into an observer canonical form. There is, however, a strong dependence on the initial conditions when using this approach. Moreover, all the above-mentioned modulating function-based observers do not take into account model disturbances.  Recently, in our previous work \cite{Yasmine2023}, an extension of the MFBM to triangular nonlinear systems subject to model disturbances was proposed to jointly estimate the states and the disturbance. \\ 

\noindent
In the present article, we propose a robust non-asymptotic observer based on modulating functions by extending the results of our previous work in \cite{Yasmine2023} to a larger class of nonlinear systems. The class of systems considered in this work is the triangular nonlinear system with multi-non-linearities, where the last equation is affected by model disturbances that encompass model uncertainties and external disturbances. The proposed observer estimates the states and the disturbance term in a step-by-step approach. The proposed modulating function-based estimator will consider both offline and online frameworks.\\

\noindent
The present article is organized as follows. Section \ref{ModulatinFunction} introduces the modulating function-based method as well as its main properties. Section \ref{estimation} presents the class of systems considered for the estimation as well as the main results for the non-asymptotic estimation of the state and disturbance. The applicability of the proposed robust modulating function estimator is illustrated in section \ref{simulation} through numerical simulations. It is first applied to an academic example, then compared to the second-order sliding mode super twisting observer under different levels of measurement noise to assess its robustness. Finally, concluding remarks and future work directions are given in section \ref{conclusion}


\section{Modulating Function Based Estimation Method}\label{ModulatinFunction}
\noindent

\noindent In this Section, the definition of modulating functions is provided as well as its main property. The offline and online estimation schemes of the modulation function-based method are also provided.

\subsection{Definition and property}
\begin{defi}\cite{Abeer2015}\label{defi1} \textit{(Modulating Function)}
A non-zero function $\phi(t): [a, b] \rightarrow \mathbb{R}$ is said to be a modulating function of order $k$, with $k \in \mathbb{N}^*$, if it satisfies the following \\
(P1): $\phi \in \mathcal{C}^k([a, b])$ \\
(P2):   $\phi^{(i)}(a)=  \phi^{(i)}(b)=0, \quad i=0,1, \ldots, k-1.$
\end{defi}

\begin{defi} \cite{Abeer2015} \textit{(Modulation operator)} \label{def2}
The modulation operator associated to the modulating function $\phi(t) \in C^{k}([a, b])$ applied to an  integrable signal $y : [a, b] \subset\mathbb{R}^{+} \rightarrow \mathbb{R}$ is given by the following inner product over the interval $I=[a, b]$:
\begin{equation*}
  \langle\phi, y\rangle_{I} =  \int_{a}^{b} \phi(t) y(t) \mathrm{d}t.
\end{equation*}
\end{defi}

\begin{proper} 
\label{property1}
The main property of the modulating function is derived using integration by parts and the boundary conditions (P2) in Definition \ref{defi1}
\begin{equation}\label{eq2}
\displaystyle\langle\phi, y^{(i)}\rangle_I=\int_a^b \phi(t) y(t)^{(i)} \mathrm{d} t
=(-1)^i \int_a^b \phi(t)^{(i)} y(t) \mathrm{d}t.
\end{equation}
\end{proper}

\vspace{0.1cm}
\noindent
In several dynamical systems, the derivatives of the output are rarely measured. However, their knowledge is necessary to get an idea about the state of the system. The Modulating Function-based method (MFBM) consists of multiplying both sides of the model differential equation by a set of modulating functions, allowing to shift the derivatives from the unknown, and possible noisy signal to the smooth modulating function which allows avoiding the amplification of noise.  Additionally, the use of the integral operator  will mitigate the effect of noise on the estimation. One of the main advantages provided by the MFBM is that the estimation of the variable of interest involves transforming the ordinary differential equations into a set of algebraic equations, therefore, the direct problem does not have to be solved, and the initial condition is no longer needed. 

\subsection{Offline and online settings}
\noindent
MFBM allows estimating the variable or parameter of interest both offline and online. The offline estimation is achieved by applying the modulation operator over the interval $I=[0; T]$ where $T$ is the final time. 
\begin{equation}\label{eq3}
\displaystyle\langle\phi, y^{(i)}\rangle_I=\int_0^T\!\!\phi(\tau) y(\tau)^{(i)} \mathrm{d} \tau=(-1)^{(i)}\int_0^T\!\!\phi^{(i)}(\tau) y(\tau) \mathrm{d} \tau.
\end{equation}
The online estimation, on the other hand, is obtained by using a sliding integration window  \cite{Liu2014}
 \begin{align}\label{eq4}
\displaystyle\langle\phi, y^{(i)}\rangle_I &=\int_{t-h}^t \phi(\tau-t+h) y(\tau)^{(i)} \mathrm{d} \tau \notag \\ \displaystyle\langle\phi, y^{(i)}\rangle_I & =(-1)^{(i)}\int_{t-h}^t\!\!\phi^{(i)}(\tau-t+h) y(\tau) \mathrm{d} \tau.
\end{align}
The modulation operator is applied over the interval $I=[t-\tau;t]$, $\forall t\in [h,T]$, and $h\in [0, T]$ representing the sliding integration window.

\noindent
In the rest of the paper, the modulation operator $\langle\phi, y^{(i)}\rangle_I$ represents, without loss of generality, both the offline and online estimation schemes. However, in the numerical simulation, the offline and online estimation setups will be distinguished and analyzed separately.

\section{Non-asymptotic State and Disturbance Estimation}
\label{estimation}
\subsection{Problem Formulation}
\noindent
We consider the following nonlinear system in the triangular canonical form with multi non-linearities
\begin{align}\label{model}
    \left\{
    \begin{array}{ll}
       &\!\!\!\dot{x} =\left[\begin{array}{cc}
       \dot{x}_1\\
       \vdots\\
       \dot{x}_{n-1}\\
       \dot{x}_n
       \end{array}\right]=\left[\begin{array}{ccccc}
              {x}_2+f_1(x_1,u)\\
              \vdots\\
              {x}_{n}+f_{n-1}(x_1,...,x_{n-1},u)\\
             f_n(x,u) + d(t)
         \end{array}\right]~\\
      &y= x_1,
    \end{array}\right.
\end{align}

\noindent
where $x\in\mathbb{R}^n$ is the system's state vector, $u\in
\mathbb{R}$ is the input bounded signal, and $y\in \mathbb{R}$ is the measured output. $f_i(x_1,...,x_i,u)$, $i=1,2,..,n$ are continuous  nonlinear functions.  $d(t)$ is a bounded unknown disturbance term that comprises model uncertainties and external disturbances.  The functions $f_i$ are assumed to be locally Lipschitz on their arguments for all bounded $x$ and $u$.\\

\noindent
The goal is to estimate the states $x_k$, $k=2,...,n$, and the disturbance term $d(t)$ using the Modulating Function-Based Method (MFBM). The triangular structure of system \ref{model}, when combined with the modulation operator and property \ref{property1}, offers the advantage to decouple the estimation of the states from the estimation of the disturbance $d(t)$. Assuming that $u(t)$ and $y(t)$ are persistently exciting, the states and the disturbance can be estimated using the MFBM in a two-step framework. First, given the input $u$ and measured output $y=x_1$, the states $x_k$, $k=2,...,n$ are estimated using the first $n-1$ equations of \eqref{model}. Once the states are estimated, the disturbance term can be estimated using the last equation of system~\ref{model}.\\

\noindent 
The estimation of the disturbance term $d(t)$ is crucial when designing controllers, as it affects the performance of the closed-loop system. In addition, estimating the disturbance provides transient performance guarantees, which is one of the main challenges in adaptive control.

\subsection{Non-asymptotic state estimation}
\noindent
Since $x_k$, $k=2,...,n$ are time-varying, we decompose each state in the space spanned by a set of unknown coefficients $a_{j,k}$ and known basis functions $\alpha_{j,k}(t)$ as follows
\begin{equation}
x_{k}(t) =\sum_{j=1}^{+\infty} a_{j,k} \alpha_{j,k}(t), \quad \forall k=2,...,n, 
\end{equation}
We usually truncate the decomposition to the first $M_k$ terms, for $k=2,...,n$.
\begin{equation}
x_{k}(t) \approx \sum_{j=1}^{M_k} a_{j,k} \alpha_{j,k}(t), \quad \forall k=2,...,n, 
\label{eq6}
\end{equation}
Indeed, each state $x_k$ is a linear combination of $M_k$ basis functions where the coefficient $a_{j,k}$ of these basis functions are estimated using the MFBM, which leads to the following proposition.

\begin{prop}
Let $\alpha_{j,k}(t)$ be the known basis functions and $\hat{a}_{j,k}$ the corresponding unknown coefficients $\forall k=2,...,n$ and let $u(t)$ $y(t)$ be the input and the output of the nonlinear system defined in \eqref{model}, respectively. Consider a set of modulating function $\{\phi_i\}_{i=1}^{i=S}$ of order $l\geqslant 1$ satisfying (P1) and  (P2). Then, the  estimation of the coefficients $a_{j,k}$ is given by the following closed-form solution 
\begin{equation}\label{EqCoefState}
\!\!\!\!\left[\!\!\begin{array}{c}
\hat{a}_{1,k}\\
\vdots \\
\hat{a}_{M_k,k}
\end{array}\!\!\right]\!\!=\!\!  -\Theta_k^{-1} 
\left[\!\!\begin{array}{c}
\langle \dot{\phi}_1, \hat{x}_{k-1} \rangle +\left\langle\phi_1, f_{k-1} (y,..,\hat{x}_{k-1},u)\right\rangle_I\\
\vdots\\
\langle \dot{\phi}_S, \hat{x}_{k-1} \rangle+ \left\langle\phi_S, f_{k-1}(y,..,\hat{x}_{k-1},u)\right\rangle_I
\end{array}\!\!\right]
\end{equation}
where
\begin{equation*}
    \Theta_k=\left[\begin{array}{c}
  \langle \phi_1(\tau), \alpha_{1}(\tau)\rangle_I \cdots \langle \phi_1(\tau), \alpha_{M}(\tau) \rangle_I\\
  \vdots\\
  \langle \phi_S(\tau), \alpha_{1}(\tau) \rangle_I\cdots \langle\phi_S(\tau), \alpha_{M}(\tau)\rangle_I
\end{array} \right],
\end{equation*}
and $S$, $M_k \in \mathbb{N}^*$ with $S\geqslant M_k$. 
\end{prop}

\paragraph{\textbf{proof}}
We start by multiplying the first $(n-1)$ equations of \eqref{model}  with the modulating function $\phi$
\begin{align}\label{pr1a}
    \left\{
    \begin{array}{ll}
      \phi \dot{x}_1(t) =& \phi {x}_2(t) + \phi f_1(x_1(t) ,u(t))\\
       \phi \dot{x}_2(t)  =& \phi {x}_3(t) +\phi f_2(x_1(t) ,x_2(t) ,u(t) )\\
       \quad \vdots & \vdots\\
       \phi \dot{x}_{n-1}(t) =&   \phi {x}_{n}(t) +\phi f_{n-1}(x_1(t) ,...,x_{n-1}(t),u(t))\\
    \end{array}\right.
\end{align}
Applying the modulation operator, one obtains 
\begin{align}
    \left\{
    \begin{array}{ll}
      \langle\phi, \dot{x}_1 \rangle_I=& \langle\phi, {x}_2\rangle_I+ \langle \phi, f_1(x_1,u)\rangle_I\\
    \langle \phi ,\dot{x}_2\rangle_I =& \langle\phi, {x}_3\rangle_I+\langle\phi ,f_2(x_1,x_2,u)\rangle_I\\
       \quad \vdots & \quad\vdots\\
      \langle\phi, \dot{x}_{n-1}\rangle_I= &  \langle\phi, {x}_{n}\rangle_I+\langle\phi, f_{n-1}(x_1,...,x_{n-1},u)\rangle_I\\
    \end{array}\right.
    \label{eq8}
\end{align}
Using Property \ref{property1} shifts the derivative from $\dot{x}_k$, $\forall k=1,..,n-1$ to the modulating function $\phi$
\begin{align}
    \left\{
    \begin{array}{ll}
      -\langle\dot{\phi}, x_1 \rangle_I=& \langle\phi, {x}_2\rangle_I+ \langle \phi, f_1(x_1,u)\rangle_I\\
    -\langle \dot{\phi} ,x_2\rangle_I =& \langle\phi, {x}_3\rangle_I+\langle\phi ,f_2(x_1,x_2,u)\rangle_I\\
       \qquad \vdots & \quad\vdots\\
      - \langle\dot{\phi}, x_{n-1}\rangle_I= &  \langle\phi, {x}_{n}\rangle_I+\langle\phi, f_{n-1}(x_1,...,x_{n-1},u)\rangle_I\\
    \end{array}\right.
    \label{eq9}
\end{align}
Since $x_k$ $, \forall k=2, ...,n$ is time-varying, it is decomposed into a space spanned by $M_k$ chosen basis functions $\alpha_{j,k}(t)$ multiplied by unknown coefficients $a_{j,k}$ as in equation \eqref{eq6}, and then substituted  into \eqref{eq9}
\begin{align}
\langle \phi, x_{k}\rangle_I &=\sum_{j=1}^{M_k} a_{j,k} \langle\phi, \alpha_{j,k}\rangle_I \notag\\
&=- \langle\dot{\phi}, x_{k-1}\rangle_I-\langle\phi, f_{k-1}(x_1,...,x_{k-1},u)\rangle_I,
\end{align}
which is equivalent in vector notations to 
\begin{align}\label{eqqa1}
\big[\langle \phi, \alpha_{1,k}\rangle_I  \hdots \langle \phi, \alpha_{M_k,k}\rangle_I \big]\left[\begin{array}{c}
a_{1,k} \\
\vdots \\
a_{M_k,k}
\end{array}\right]=&-\langle \dot{\phi}, x_{k-1}\rangle_I -\langle\phi, f_{k-1}\rangle_I.
\end{align}
To estimate the coefficients $a_{j,k}, j=1, \ldots, M_k$, at least $M_k$ linearly independent equations are required. To that end,  $S\geqslant M_k$ different modulating functions $\phi_i$, $i=1, \ldots, S$ of order $l\geqslant 1$ are used, which leads to the following algebraic system
\begin{equation}
\label{SystemState}
\Theta_k\!\!\left[\!\!\begin{array}{c}
\hat{a}_{1,k}\\
\vdots \\
\hat{a}_{M_k,k}
\end{array}\!\!\right]\!\!=\!\! -
\!\!\left[\!\!\begin{array}{c}
\langle \dot{\phi}_1, \hat{x}_{k-1} \rangle +\left\langle\phi_1, f_{k-1} (y,..,\hat{x}_{k-1},u)\right\rangle_I\\
\vdots\\
\langle \dot{\phi}_S, \hat{x}_{k-1} \rangle+ \left\langle\phi_S, f_{k-1}(y,..,\hat{x}_{k-1},u)\right\rangle_I
\end{array}\!\!\right]
\end{equation}
where $\hat{a}_{j,k}$ are estimates of $a_{j,k}$, the states $x_2, ..., x_{k-1}$ have been substituted by their estimates  $\hat{x}_2, ..., \hat{x}_{k-1}$, for $k=2,...,n$,  and the output $y$ was used instead of the state $x_1$ . $\Theta_k$ is an $S\times M_k$ matrix given by

\begin{equation*}
    \Theta_k=\left[\begin{array}{c}
  \langle \phi_1(\tau), \alpha_{1,k}(\tau)\rangle_I \cdots \langle \phi_1(\tau), \alpha_{M_k,k}(\tau) \rangle_I\\
  \vdots\\
  \langle \phi_S(\tau), \alpha_{1,k}(\tau) \rangle_I\cdots \langle\phi_S(\tau), \alpha_{M_k,k}(\tau)\rangle_I
\end{array} \right].
\end{equation*}
Finally, the parameters $\hat{a}_{j,k}$, $\forall j=1, \ldots, M_k$, are obtained by solving \eqref{SystemState}, and the estimated states $\hat{x}_k$ is given by 
\begin{equation}
\hat{x}_{k}(t) =\sum_{j=1}^{M_k} \hat{a}_{j,k} \alpha_{j,k}(t), \quad \forall k=2,...,n
\end{equation}

\begin{lemma}
The algebraic system \eqref{EqCoefState} is equivalent to the following system for a set of modulating functions $\{\phi_i\}_{i=1}^{i=S}$ of order $l\geqslant k$ satisfying (P1) and  (P2), for $k=2,...,n$
\begin{equation}
\left[\!\!\begin{array}{c}
\hat{a}_{1,k}\\
\vdots \\
\hat{a}_{M_k,k}
\end{array}\!\!\right]\!\!= \Theta_k^{-1} 
\Psi_k
\end{equation}
where 
\begin{equation}
   \Psi_k= \left[\psi_k^1 \hspace*{0.2cm}\psi_k^2 \hspace*{0.2cm}... \hspace*{0.2cm}\psi_k^S\right]^T
\end{equation} 
with $\psi_k^i$ for $k=2,...,n$ and $i=1,...,S$ given by
\begin{equation}
\psi_k^i= (-1)^{(k-1)}\langle {\phi_i}^{(k-1)}, y \rangle_I-\sum_{r=1}^{k-1}(-1)^{(k-1-r)}\left\langle\phi_i^{(k-1-r)}, f_r\right\rangle_I
\end{equation}
\end{lemma}

\paragraph{\textbf{proof}}

Step 1: We start by proving by induction that 
\begin{equation}
    \dot{x}_k(t)=x_1^{(k)}(t) - \sum_{j=1}^{k-1} f_i^{(k-i)}(x_1(t),.., x_i(t),u(t)), \forall k=2,...,n.
\end{equation}
For $k=2$, the result is straightforward. We consider the first equation of system \eqref{model} and differentiate with respect to time, which leads to the following equation
\begin{equation}
 \dot{x}_2(t)=\ddot{x}_1(t)-\dot{f}_1(x_1(t),u(t)).   
\end{equation}
Now assume that the following hold true for $k$
\begin{equation}
    \dot{x}_k(t)=x_1^{(k)}(t) - \sum_{r=1}^{k-1} f_i^{(k-r)}(x_1(t),.., x_r(t),u(t)), \forall k=2,...,n
    \label{eqind1}
\end{equation}
and prove that the above equality holds true for $k+1$.

\noindent
Considering the $(k)$-th equation of system \eqref{model} and differentiating it with respect to time 
\begin{equation}  \label{eqind2}
  \ddot{x}_{k}(t)=\dot{x}_{k+1}(t)+\dot{f}_{k}(x_1(t),...,x_{k}(t),u(t)).
\end{equation}
Differentiating \eqref{eqind1} w.r.t time
\begin{equation}
    \ddot{x}_k(t)=x_1^{(k+1)}(t) - \sum_{r=1}^{k-1} f_r^{(k+1-r)}(x_1(t),.., x_r(t),u(t)), 
    \label{eqind3}
\end{equation}
Substituting \eqref{eqind3} into \eqref{eqind2}
\begin{align}
\!\!\!\! &\dot{x}_{k+1}(t)= x_1^{(k+1)}(t) \notag\\&- \sum_{r=1}^{k-1} f_r^{(k+1-r)}(x_1(t),.., x_r(t),u(t))-\dot{f}_{k}(x_1(t),...,x_{k}(t),u(t))  
\end{align}
Finally, one obtains
\begin{equation}
   \dot{x}_{k+1}(t)= x_1^{(k+1)}(t) - \sum_{r=1}^{k} f_r^{(k+1-r)}(x_1(t),.., x_r(t),u(t))
\end{equation}
    
\noindent
Step 2: Recall now equation \eqref{eq8}
\begin{align}
    \left\{
    \begin{array}{ll}
      \langle\phi, \dot{x}_1 \rangle_I=& \langle\phi, {x}_2\rangle_I+ \langle \phi, f_1(x_1,u)\rangle_I\\
    \langle \phi ,\dot{x}_2\rangle_I =& \langle\phi, {x}_3\rangle_I+\langle\phi ,f_2(x_1,x_2,u)\rangle_I\\
       \qquad \vdots & \quad\vdots\\
      \langle\phi, \dot{x}_{n-1}\rangle_I= &  \langle\phi, {x}_{n}\rangle_I+\langle\phi, f_{n-1}(x_1,...,x_{n-1},u)\rangle_I\\
    \end{array}\right.
    \label{eq19}
\end{align}
Substituting $\dot{x}_k$, for $k=2,..,n-1$, by their expression in \eqref{eqind1} in the left-hand-side of equation \eqref{eq19}
 \begin{align*}\label{pr2}
    \left\{
    \begin{array}{ll}
      \langle \phi ,{x}_2\rangle_I &= \langle\phi, \dot{x}_1\rangle_I - \langle\phi, f_1(x_1,u)\rangle_I\\
       \langle\phi, {x}_3\rangle_I&= \langle\phi ,\ddot{x}_1\rangle_I-\langle\phi, \dot{f}_1(x_1,u)\rangle_I-\langle\phi, f_2(x_1,x_2,u)\rangle_I\\
        \langle\phi, {x}_4\rangle_I &= \langle\phi,{x}^{(3)}_1\rangle_I- \langle\phi,\ddot{f}_1(x_1,u)\rangle_I-\langle\phi,\dot{f}_2(x_1,x_2,u)\rangle_I \\ & -\langle\phi, f_3(x_1, x_2, x_3,u)\rangle_I\\
       \quad \vdots & \quad \vdots\\
       \langle\phi, {x}_{n}\rangle_I & =\langle\phi,{x}_{1}^{(n-1)}\rangle_I- \displaystyle\sum_{r=1}^{n-1}\langle\phi, {f}_{r}^{n-r-1}(x_1,...,x_{r},u)\rangle_I\\
    \end{array}\right.
\end{align*}
Applying Property \ref{property1}, one obtains for the states $x_k$, $k=2,...,n$
\begin{equation*}
  \langle\phi, x_k\rangle=  (-1)^{(k-1)}\langle {\phi}^{(k-1)}, x_1\rangle_I-\sum_{r=1}^{k-1}(-1)^{(k-1-r)}\left\langle\phi^{(k-1-r)}, f_r\right\rangle_I 
\end{equation*}
Substituting $x_k$ by its decomposition (equation \eqref{eq6}) leads to the following algebraic equation
\begin{align}\label{eq24}
\big[\langle \phi, \alpha_{1,k}\rangle_I  \hdots \langle \phi, \alpha_{M_k,k}\rangle_I \big]\left[\begin{array}{c}
a_{1}^k \\
\vdots \\
a_{M}^k
\end{array}\right]=&\psi_k
\end{align}
where
\begin{equation*}
\psi_k=(-1)^{(k-1)}\langle {\phi}^{(k-1)}, x_1\rangle_I-\sum_{r=1}^{k-1}(-1)^{(k-1-r)}\left\langle\phi^{(k-1-r)}, f_r\right\rangle_I     
\end{equation*}
To solve system \eqref{eq24}, one needs $S\geqslant M$ equations obtained by using $S$ modulating functions of order $l\geqslant k$, which leads to \begin{equation}
\label{eq25}
\Theta_k\left[\!\!\begin{array}{c}
\hat{a}_{1,k}\\
\vdots \\
\hat{a}_{M_k,k}
\end{array}\!\!\right]\!\!= 
\Psi_k
\end{equation}
where  $\hat{a}_{j,k}$ are the estimated values of the unknown coefficients $a_{j,k}$, and the vector $\Psi_k$ is given by
\begin{equation*}
   \Psi_k= \left[\psi_k^1 \hspace*{0.2cm}\psi_k^2 \hspace*{0.2cm}... \hspace*{0.2cm}\psi_k^S\right]^T
\end{equation*} 
\noindent
with $\psi_k^i$ for $k=2,...,n$ and $i=1,...,S$ given by, where $x_1$ is substituted by $y$, and the states $x_2,...,x_r$ by their estimated values $\hat{x}_2,..., \hat{x}_r$, for $r=1,...,k-1$, and $k=2,...,n$.
\begin{equation*}
\psi_k^i= (-1)^{(k-1)}\langle {\phi_i}^{(k-1)}, y \rangle_I-\sum_{r=1}^{k-1}(-1)^{(k-1-r)}\left\langle\phi_i^{(k-1-r)}, f_r\right\rangle_I
\end{equation*}
Therefore the coefficient $\hat{a}_{j,k}$, for $j=1,..,M$, $k=2,...,n$ are obtained by solving either system \eqref{SystemState} or system \eqref{eq25}, which concludes the proof.

\subsection{Disturbance estimation }
\noindent
In most cases, the disturbance $d(t)$ is time-varying and can be in the form of model uncertainties or external disturbances. Following the same reasoning as the state estimation in the previous subsection, the disturbance $d(t)$ can be decomposed into a sum of known basis functions $\beta_{j}(t)$ and unknown coefficients $b_{j}$, which is truncated to the first $N$ terms
\begin{equation}
d(t) =  \sum_{j=1}^{\infty} b_{j}\beta_{j}(t)\approx\sum_{j=1}^{N} b_{j}\beta_{j}
\label{eq28}
\end{equation}

\noindent
 Given the estimated states $\hat{x}_{k}$, $ \forall k=2,...,n$ and the measured input $u(t)$ and output $y(t) = x_{1}(t)$, The estimation of the disturbance term $d(t)$ using the MFBM can be achieved using the last equation of system \eqref{model}. The estimation of the disturbance is given in the following proposition. 

\begin{prop} 
Let $u(t)$ and $y(t)=x_{1}(t)$ be respectively  the measured input and output of system \eqref{model}, and  $\hat{x}_{k}$ the estimated states $ \forall k=2,...,n$. Let $\{\phi_i\}_{i=1}^{i=D}$ be a set of modulating functions of order $l\geqslant 1$ satisfying (P1) and (P2). Then an estimate of the disturbance term $d(t)$ is given by
\begin{equation}
 \hat{d}(t) =  \sum_{j=1}^{N} \hat b_{j}\beta_{j}(t)
 \label{eq34}
\end{equation}
where $N, D \in \mathbb{N}^{*}$, with $D\geqslant N$, $\beta_{j}(t)$ are  known basis functions and $\hat b_{j}$ are estimates of the unknown coefficients given by the following closed-form solution
\begin{equation}
  \begin{bmatrix}
\hat b_{1}\\
\vdots \\
\hat b_{N}
\end{bmatrix} = 
- \Theta_{d}^{-1} 
  \begin{bmatrix}
\langle\dot \phi_1, \hat{x}_n\rangle_I + \langle\phi_1, f_n(y,..,\hat{x}_n,u)\rangle_I \\
\vdots \\
\langle\dot \phi_D, \hat{x}_{n}\rangle_I + \langle\phi_D, f_n(y,..,\hat{x}_n,u)\rangle_I
\label{coeffd}
\end{bmatrix}
\end{equation}
where $\Phi_{d}$ is given by
\begin{equation}\label{eqq1}
\Theta_{d} = 
\begin{bmatrix}
\langle\phi _1, \beta_{1}(t)\rangle_I & \hdots & \langle\phi _1, \beta_{N}(t)\rangle_I \\
\vdots & \ddots & \vdots \\
\langle\phi _D, \beta_{1}(t)\rangle_I & \hdots & \langle\phi _D, \beta_{N}(t)\rangle_I
\end{bmatrix}
\end{equation}
\end{prop}

\vspace*{0.5cm}

\paragraph{\textbf{proof}}
Considering the last equation of system \eqref{model} and applying the modulation operator
\begin{equation}
    \langle\phi, \dot{x}_{n}\rangle_I=   \langle\phi, {f}_{n}(x,u)\rangle_I+\langle\phi, d(t)\rangle_I\\
\end{equation}

\noindent
Using Property \ref{property1}, the derivative is shifted from $x_n$ to the modulating function
\begin{equation}
    -\langle\dot{\phi}, x_{n}\rangle_I=   \langle\phi, {f}_{n}(x,u)\rangle_I+\langle\phi, d(t)\rangle_I
    \label{eq32}
\end{equation}
Given that the disturbance term $d(t)$ is time-varying, it is decomposed into the space spanned by a set of  known basis functions $\beta_{j}(t)$ multiplied by unknown coefficients $b_{j}$, for $j=1,..., N$, as given by equation \eqref{eq28}. \\
Rearranging \eqref{eq32}, and substituting $d$ by \eqref{eq28} 
\begin{equation}
\langle \phi, d\rangle_I =\sum_{j=1}^{N} b_{j} \langle\phi, \beta_{j}\rangle_I \notag\\
=- \langle\dot{\phi}, x_{n}\rangle_I-\langle\phi, f_{n}(x,u)\rangle_I,
\end{equation}
which is equivalent to 
\begin{align}\label{eqqb1}
\big[\langle \phi, \beta_{1} \rangle_I  \hdots \langle \phi, \beta_{N} \rangle_I \big]\left[\begin{array}{c}
\hat{b}_{1} \\
\vdots \\
\hat{b}_{N}
\end{array}\right]=&-\langle \dot{\phi}, \hat{x}_{n}\rangle_I -\langle\phi, f_{n}(\hat{x},u)\rangle_I.
\end{align}

\noindent
Where $\hat{b}_j$ are the estimated values of the unknown coefficients $\hat{b}_j$, $j=1,..N$, and the states $x_k$ are substituted by their estimates $\hat x_k$, $\forall k=2,...,n$, and the $x_1$ is substituted by the measured output $y$.\\

\noindent
To solve for $\hat{b}_j$, $j=1,..N$, we need $D$ equations, where $D\geqslant N$, which leads to the following system of algebraic equation 
\begin{equation}\label{solv_d}
 \Theta_{d}  \begin{bmatrix}
\hat b_{1}\\
\vdots \\
\hat b_{N}
\end{bmatrix} = 
- 
  \begin{bmatrix}
\langle\dot \phi_1, \hat{x}_n\rangle_I + \langle\phi_1, f_n(\hat{x},u)\rangle_I \\
\vdots \\
\langle\dot \phi_D, \hat{x}_{n}\rangle_I + \langle\phi_D, f_n(\hat{x},u)\rangle_I
\end{bmatrix}
\end{equation}
where $\Theta_{d}$ is a $D\times N$  matrix
\begin{equation}
\Theta_{d} = 
\begin{bmatrix}
\langle\phi _1, \beta_{1}(t)\rangle_I & \hdots & \langle\phi _1, \beta_{N}(t)\rangle_I \\
\vdots & \ddots & \vdots \\
\langle\phi _D, \beta_{1}(t)\rangle_I & \hdots & \langle\phi _D, \beta_{N}(t)\rangle_I
\end{bmatrix}
\end{equation}
Finally, the coefficients $\hat{b}_j$, for $j=1,...,N$ are obtained by solving~\eqref{solv_d}, and the estimated disturbance is given by 
\begin{equation}
 \hat{d}(t) =  \sum_{j=1}^{N} \hat b_{j}\beta_{j}(t)
\end{equation}

\begin{lemma}
For a set of modulating functions $\{\phi_i\}_{i=1}^{i=D}$of order $l\geqslant n$ satisfying (P1) and  (P2), the algebraic system \eqref{coeffd} is equivalent to the following system
\begin{equation*}
  \begin{bmatrix}
\hat b_{1}\\
\vdots \\
\hat b_{N}
\end{bmatrix} = 
 \Theta_{d}^{-1} 
  \begin{bmatrix}
(-1)^n\langle \dot\phi_1, y\rangle_I -\displaystyle\sum_{j=1}^{n}(-1)^{(n-i)}\left\langle\phi_1^{(n-i)}, f_i\right\rangle_I \\
\vdots \\
(-1)^n \langle \dot\phi_D, y\rangle_I -\displaystyle\sum_{j=1}^{n}(-1)^{(n-i)}\left\langle\phi_D^{(n-i)}, f_i\right\rangle_I 
\label{discoeff}
\end{bmatrix}
\end{equation*}
Then, the estimate of the disturbance is given as follows
\begin{equation*}
 \hat{d}(t) =  \sum_{j=1}^{N} \hat b_{j}\beta_{j}(t).
\end{equation*}
\end{lemma}

\paragraph{\textbf{proof}}
Considering the last equation of system \eqref{model}
\begin{equation}
   \dot{x}_n=f_n(\hat{x},u)+d(t),
   \label{eq44}
\end{equation}
Recall that
\begin{equation}
    \dot{x}_n=x_1^{(n)} - \sum_{r=1}^{n-1} f_r^{(n-r)}(x_1,.., x_r,u).
\end{equation}
Equation \eqref{eq44} becomes
\begin{equation}
    d(t)= x_1^{(n)} - \sum_{r=1}^{n} f_r^{(n-r)}(x_1,.., x_r,u).
\end{equation}
Applying the modulation operator and Property \ref{property1}
\begin{equation}
\langle \phi, d\rangle_I = (-1)^{n}\langle\phi^{(n)}, x_{1}\rangle_I- \sum_{r=1}^{n} (-1)^{n-r} \langle\phi^{(n-r)}, f_{r}(x,u)\rangle_I,
\label{eq47}
\end{equation}
Substituting \eqref{eq34} into \eqref{eq47} and writing it in a vector form, one obtains 
\begin{align*}
\big[\langle \phi, \beta_{1} \rangle_I  \hdots \langle \phi, \beta_{N} \rangle_I \big]\left[\begin{array}{c}
\hat{b}_{1} \\
\vdots \\
\hat{b}_{N}
\end{array}\right]&=(-1)^{n}\langle\phi^{(n)}, y\rangle_I\\ \notag
&- \sum_{r=1}^{n} (-1)^{n-r} \langle\phi^{(n-r)}, f_{r}(x,u)\rangle_I.
\end{align*}
Where $\hat{b}_j$ are the estimated values of the unknown coefficients $\hat{b}_j$, $j=1,..N$, and the states $x_k$ are substituted by their estimates $\hat x_k$, $\forall k=2,...,n$, and the $x_1$ is substituted by the measured output $y$.\\

\noindent
To solve for $\hat{b}_j$, $j=1,..,N$, one needs $D$ equations, where $D\geq N$, obtained by using  $D$ different modulating functions of order $k\geq n$, which leads  
\begin{equation*}
   \Theta_{d}\begin{bmatrix}
\hat b_{1}\\
\vdots \\
\hat b_{N}
\end{bmatrix} = 
  \begin{bmatrix}
(-1)^n\langle \dot\phi_1, y\rangle_I -\displaystyle\sum_{j=1}^{n}(-1)^{(n-i)}\left\langle\phi_1^{(n-i)}, f_i\right\rangle_I \\
\vdots \\
(-1)^n \langle \dot\phi_D, y\rangle_I -\displaystyle\sum_{j=1}^{n}(-1)^{(n-i)}\left\langle\phi_D^{(n-i)}, f_i\right\rangle_I 
\label{discoeff1}
\end{bmatrix}
\end{equation*}
where $\Theta_{d}$ is given as \eqref{eqq1}.

\section{Numerical simulation}
\label{simulation}
\noindent
To illustrate the performance of the proposed  Modulating function Based estimation method, we consider two numerical examples. First, the proposed estimator is applied to an academic example of a third-order system. Then, the MFBM is compared to the second-order super-twisting observer \cite{Fridman2005}. Finally, a robustness analysis against the measurement noise is performed. 
\subsection{Academic example}
\noindent
Consider the following third-order nonlinear triangular system   
\begin{align}\label{sim1}
    \left\{
    \begin{array}{ll}
       &\dot{x} =\left[\begin{array}{cc}
       \dot{x}_1\\
       \dot{x}_{2}\\
       \dot{x}_3     \end{array}\right]=\left[\begin{array}{ccccc}
              {x}_2-x_1^{2}\\
              {x}_{3}-x_1 x_2\\
             \frac{-x_3^2}{1+x_1^2} + d(t)
         \end{array}\right]~\\
      &y= x_1,
    \end{array}\right.
\end{align}
In this example, a general form of the disturbance term $d(t)$ that is time and state-dependent is considered.
\begin{equation*}
d(t)=0.1\cos\left(2\frac{\pi}{5}t-\frac{\pi}{4}\right)\times(1+0.1t)\times x_1\times x_2 \times x_3.
\end{equation*}
For this example, both offline and online schemes are considered. 
The states $x_2$ and $x_3$, and  the disturbance $d(t)$ are decomposed respectively into polynomial basis functions $\alpha_{j,k}(t)= t^{j-1} $ , for $k=2, 3$,  and $\beta_{j}(t)= t^{j-1}$ as follows
\begin{equation*}
x_{2}(t) \approx \sum_{j=1}^{M_2} a_{j,2}\alpha_{j,2}(t), \quad x_{3}(t) \approx \sum_{j=1}^{M_3} a_{j,3} \alpha_{j,3}(t),
\end{equation*}

\begin{equation*}
d(t) \approx \sum_{j=1}^{N} b_{j} \beta_{j}(t).
\end{equation*}

\subsubsection{Offline estimation}
 The selected modulating functions are   normalized polynomial  modulating functions \cite{Liu2013} given by
\begin{equation}
\phi_i(t)=\frac{\bar{\phi}_i(t)}{\|\bar{\phi}_i(t)\|_{ \mathcal{L}_2}},
\end{equation}
where $\| . \|_{ \mathcal{L}_2}$ is the ${ \mathcal{L}_2}$-norm and the modulating functions $\bar{\phi}_i(t)$  for the state and disturbance estimation are given by
\begin{equation*}
  \bar{\phi}_i(t)=(T-t)^{(p_{x,k}+i)} t^{(p_{x,k}+S_k+1-i)}, \forall i=1,\cdots, S_k,  k=2,3,
  \end{equation*}
\begin{equation*}
\bar{\phi}_i(t)=(T-t)^{(p_d+i)} t^{(p_{x,k}+D+1-i)}, i=1,2, \cdots, D.
 \end{equation*}  
where $T$ is the simulation final time, $S$ and $D$ represent the number of modulating functions for the state and disturbance estimation, respectively. $ p_{x,k}, p_d\in \mathbb{N}^*$ are degrees of freedom. The parameters of the modulating functions for the simulation are given as follows $S_2=M_2=12$, $S_3=M_3=10$, and $D=N=9$. $p_{x,k}=2$, $k=2,3$ and $p_d=3$.\\
The parameter vectors $\hat{a}_2$, $\hat{a}_3$, and $\hat{b}$ obtained by the modulating function-based method for $\hat{x}_2$, $\hat{x}_3$ and $\hat{d}$, respectively, are given by  :

\begin{align*}
   \hat{a}_2=&[1.98 \text{ } -1.82 \text{ } 2.97 \text{ } -3.36 \text{ } 2.55 \text{ } -1.37  \text{ } 0.52 \text{ } -0.14\\ 
   & 0.02 \text{ } -3.26\times 10^{-3} \text{ } 2.35 \times 10^{-4}  \text{ } -7.55\times 10^{-6}]^T 
\end{align*}
\begin{align*}
   \hat{a}_3=&[2.01 \text{ } -0.27  \text{ } -0.789 \text{ }  0.92 \text{ } -0.59 \text{ } 0.24 \text{ } -0.06\\
   & 9.43\times 10^{-3} \text{ } -8.07 \times 10^{-4} \text{ } 2.95 \times 10^{-5}]^T 
\end{align*}
\begin{align*}
 \hat{b}=&[0.52 \text{ } -0.24 \text{ } 0.15 \text{ } -0.39 \text{ } 0.29 \text{ } -0.10 \text{ } 0.02 \text{ } -1.5\times 10^{-3}\\
 & 4.92 \times 10^{-5}]^T
\end{align*}

\noindent
The choice of the parameters $M_2, M_3$, and $N$ is crucial to obtain a good estimation. The bigger the coefficient the better the approximation of the states and disturbance is. However, the computation burden increases which may lead to numerical instabilities. Therefore, there is a trade-off between the accuracy of the approximation and the computation complexity. It can be noticed from the obtained parameters that the last values decrease fast to zero. Therefore, one can deduce that the choice of the truncation parameters was suitable for our problem.        

\begin{figure*}[h!]
        \centering
     \begin{overpic}[scale=0.2]{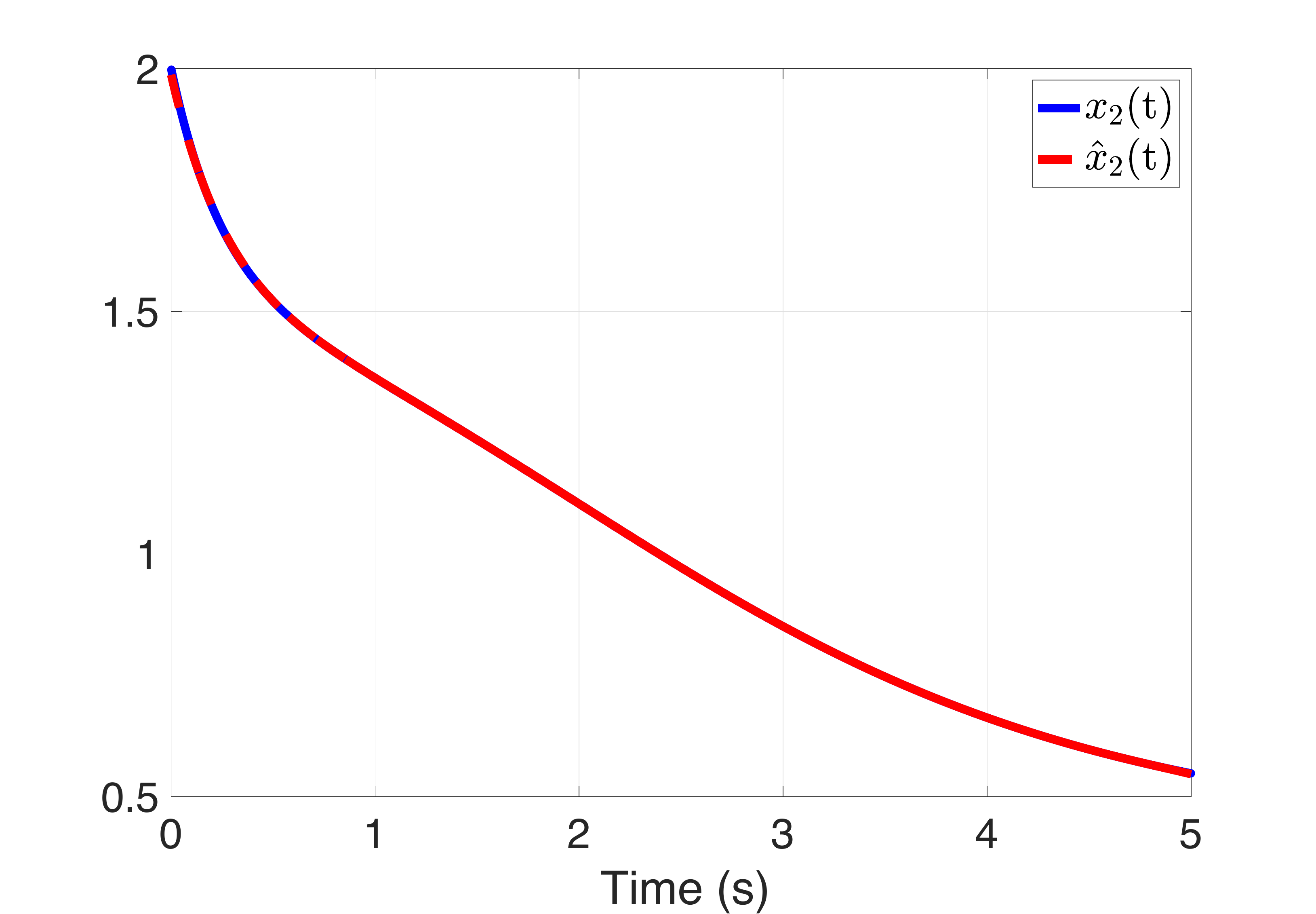}
      \put(37,42){\begin{overpic}[scale=0.09]{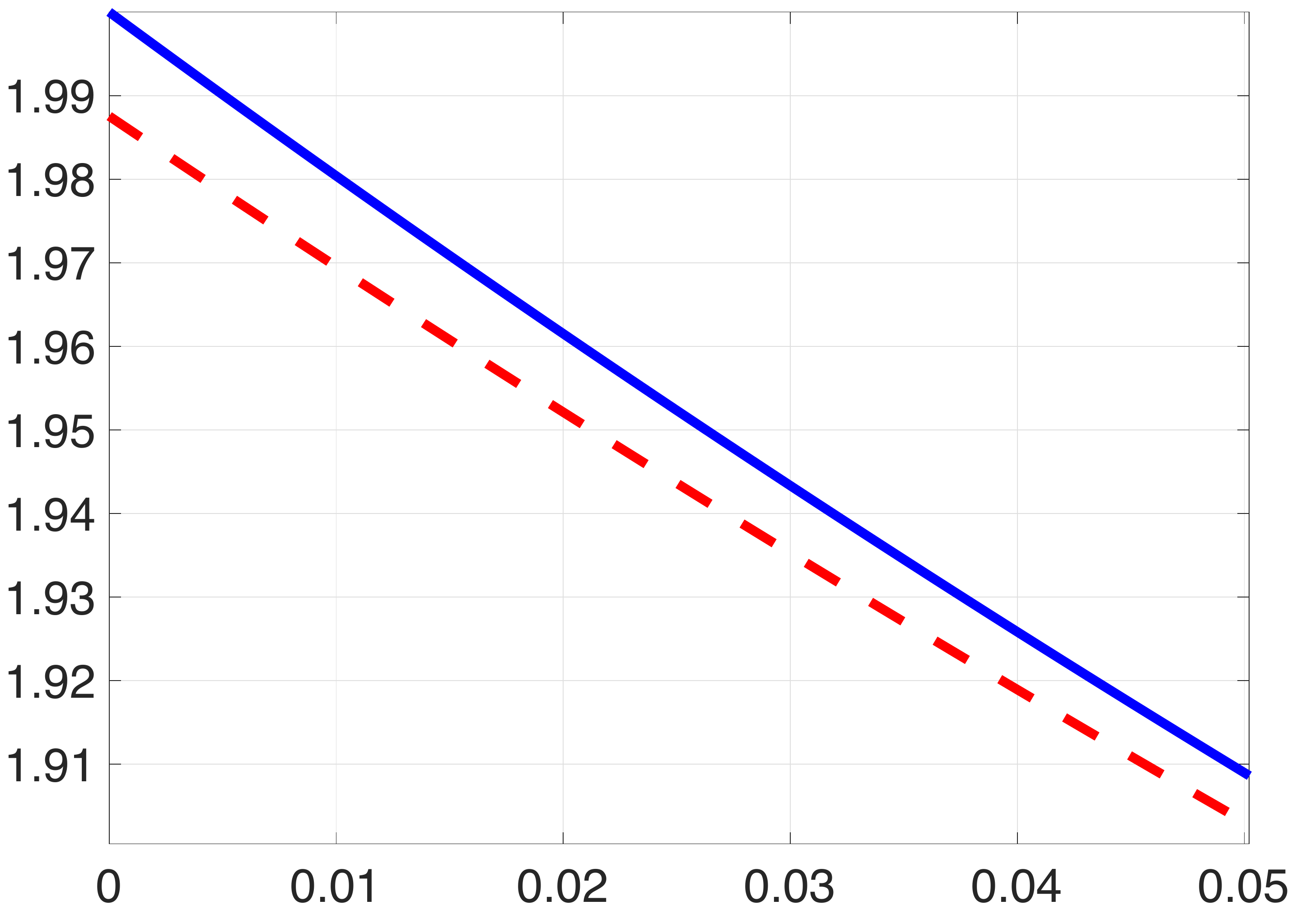} 
         \end{overpic}}
            \end{overpic}  \vspace{-2pt}
     \begin{overpic}[scale=0.2]{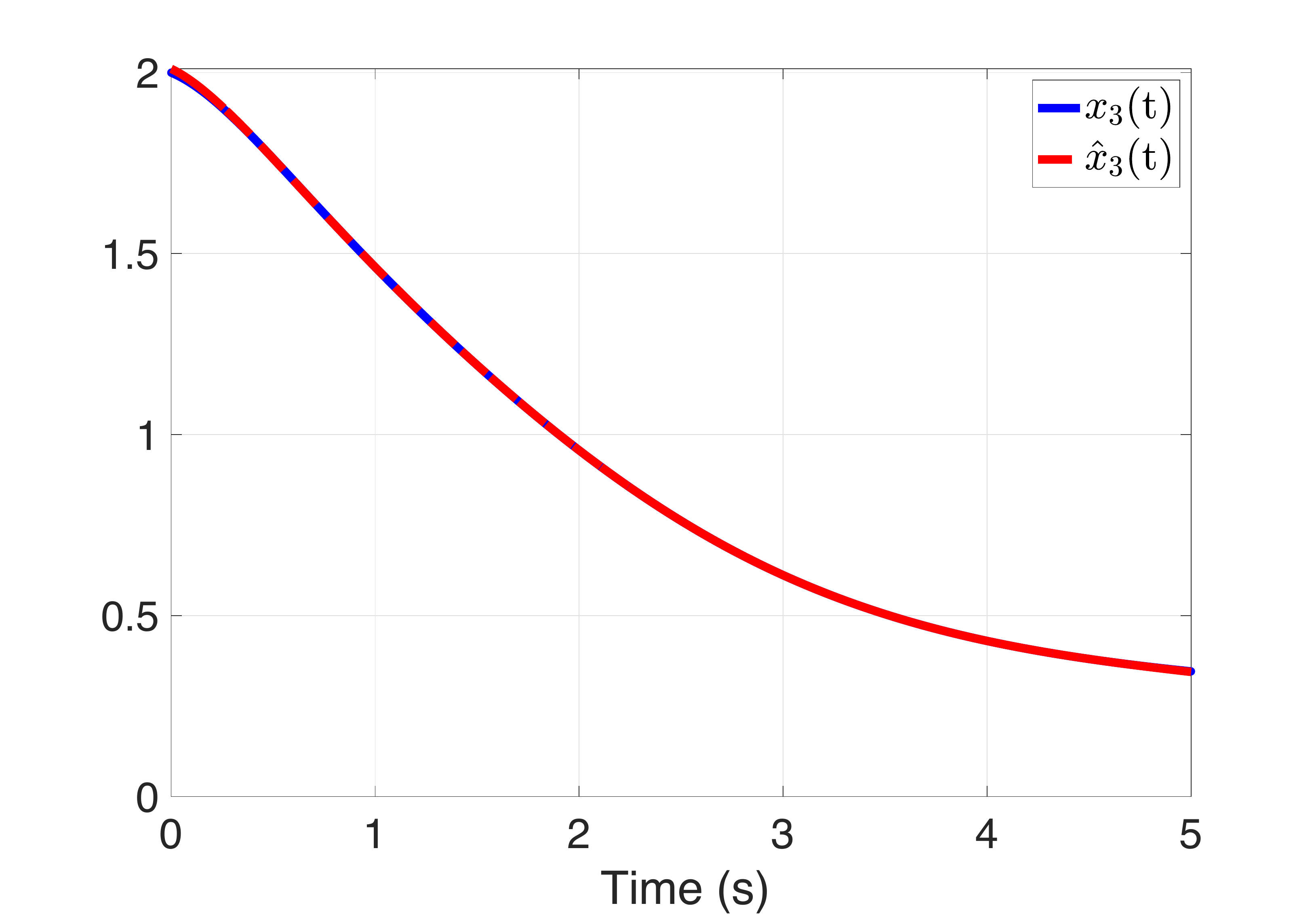}
      \put(43,44){\begin{overpic}[scale=0.075]{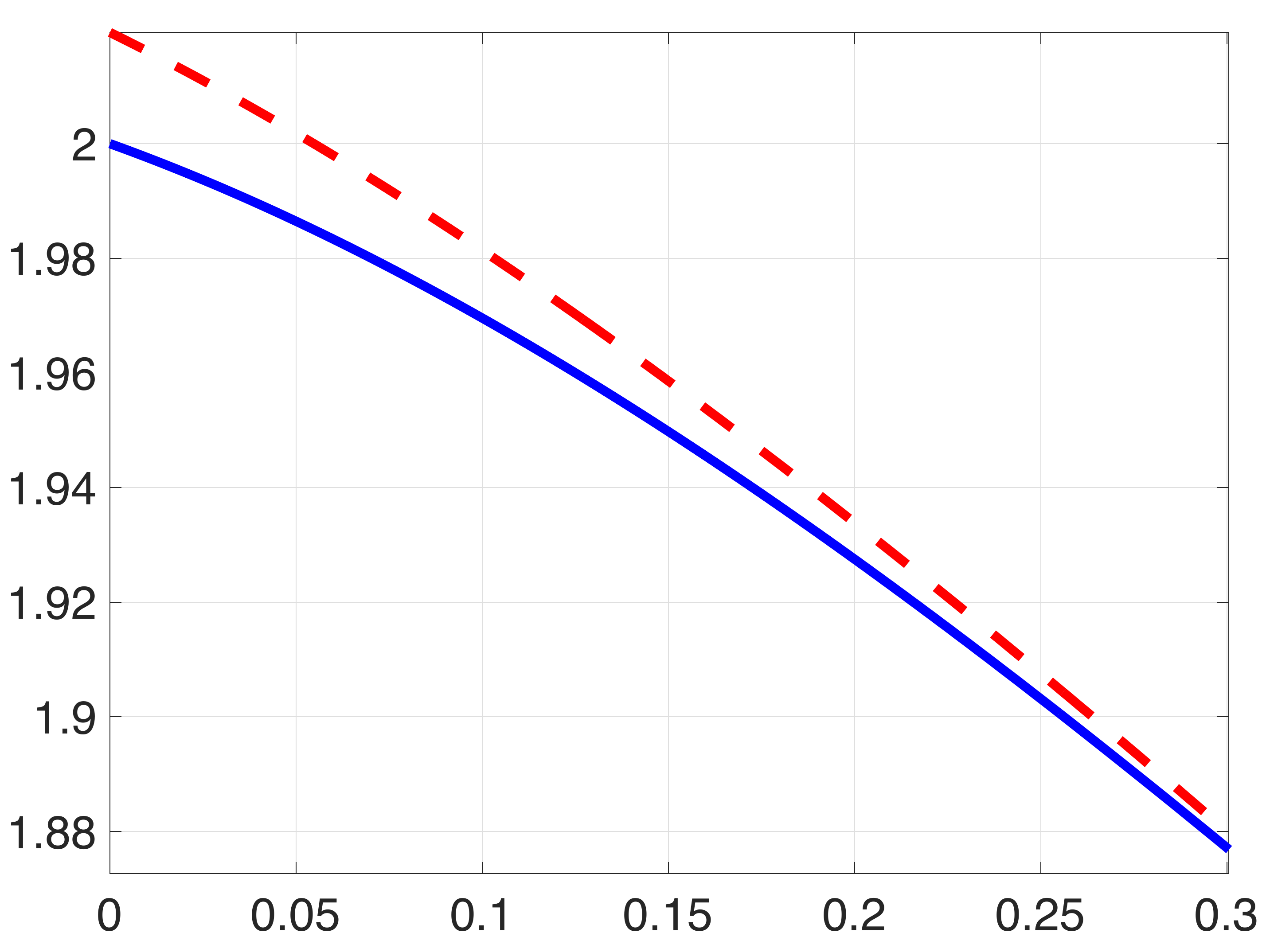} 
         \end{overpic}}
            \end{overpic}  \vspace{-2pt}
     \includegraphics[scale=0.195]{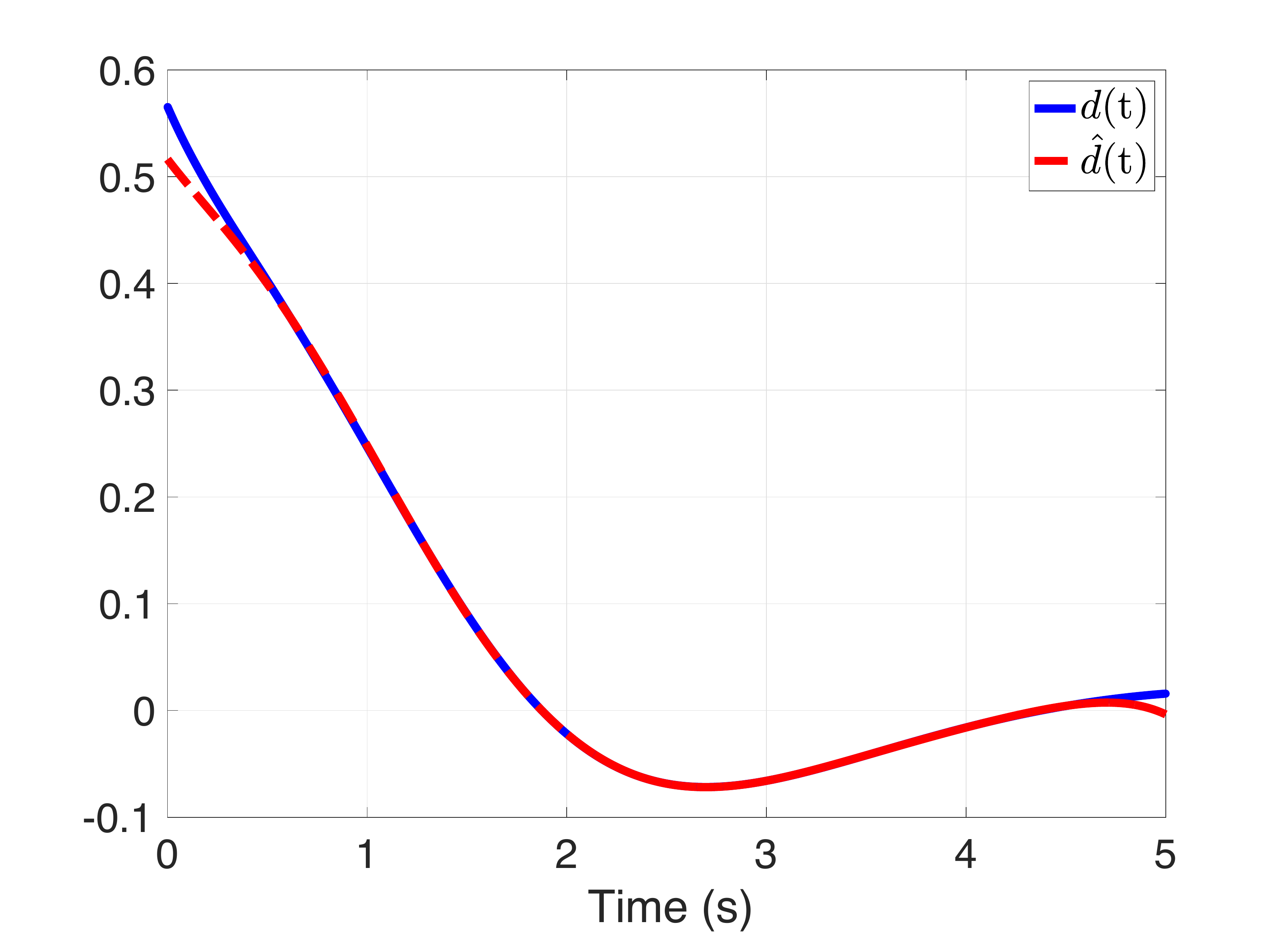}
      \caption{Offline estimation of the states and the disturbance term. }\label{fig1}
\end{figure*}

\noindent
Figure \ref{fig1} shows that the modulating function-based estimator was able to accurately estimate the states and the disturbance. However, some estimation errors at the boundary can be noticed for the estimation of $d(t)$, which is common when using a modulating function with polynomial basis functions \cite{Liu2014}. 

\begin{figure*}[!t]
        \centering
      \begin{overpic}[scale=0.2]{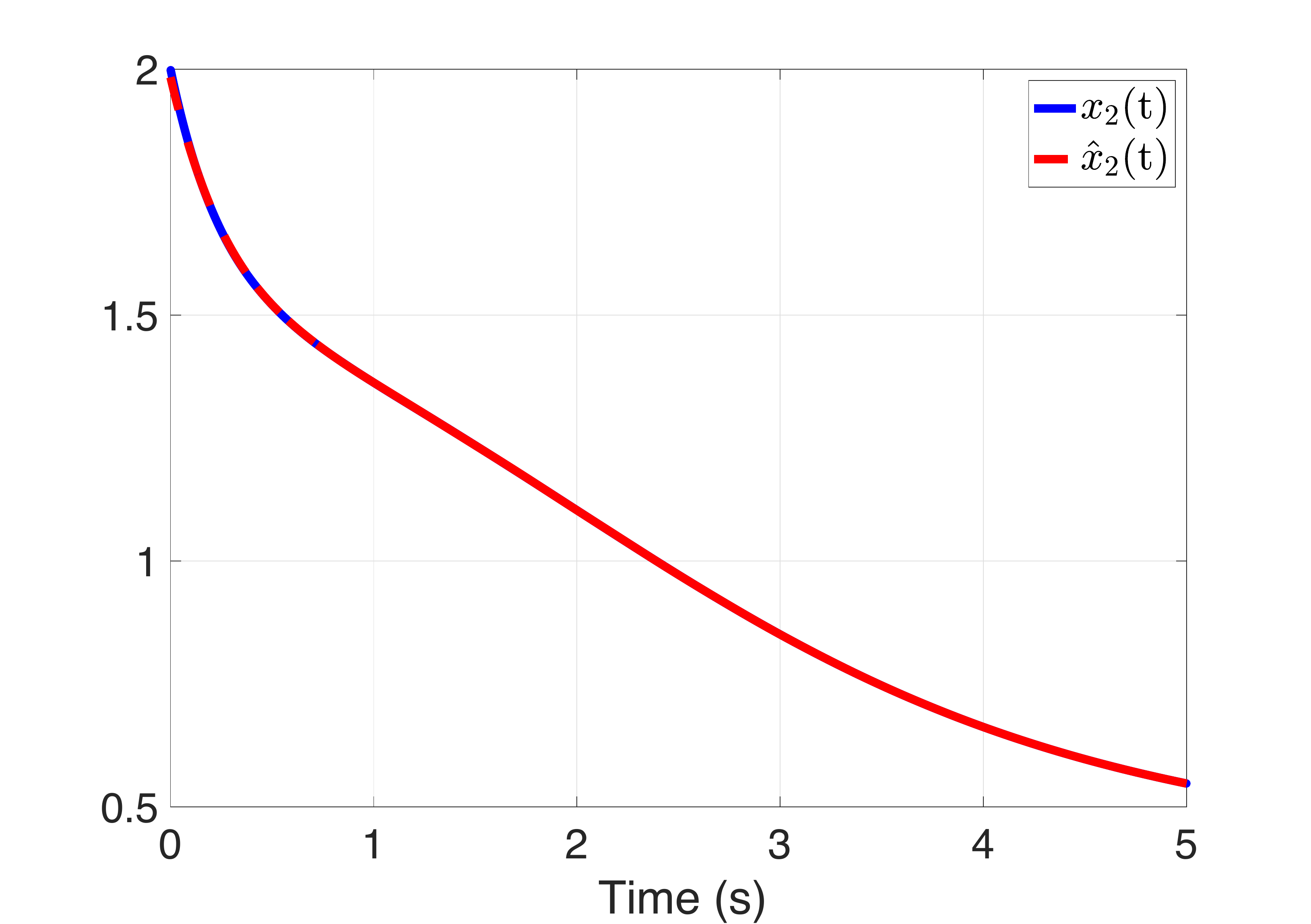}
      \put(40,44){\begin{overpic}[scale=0.08]{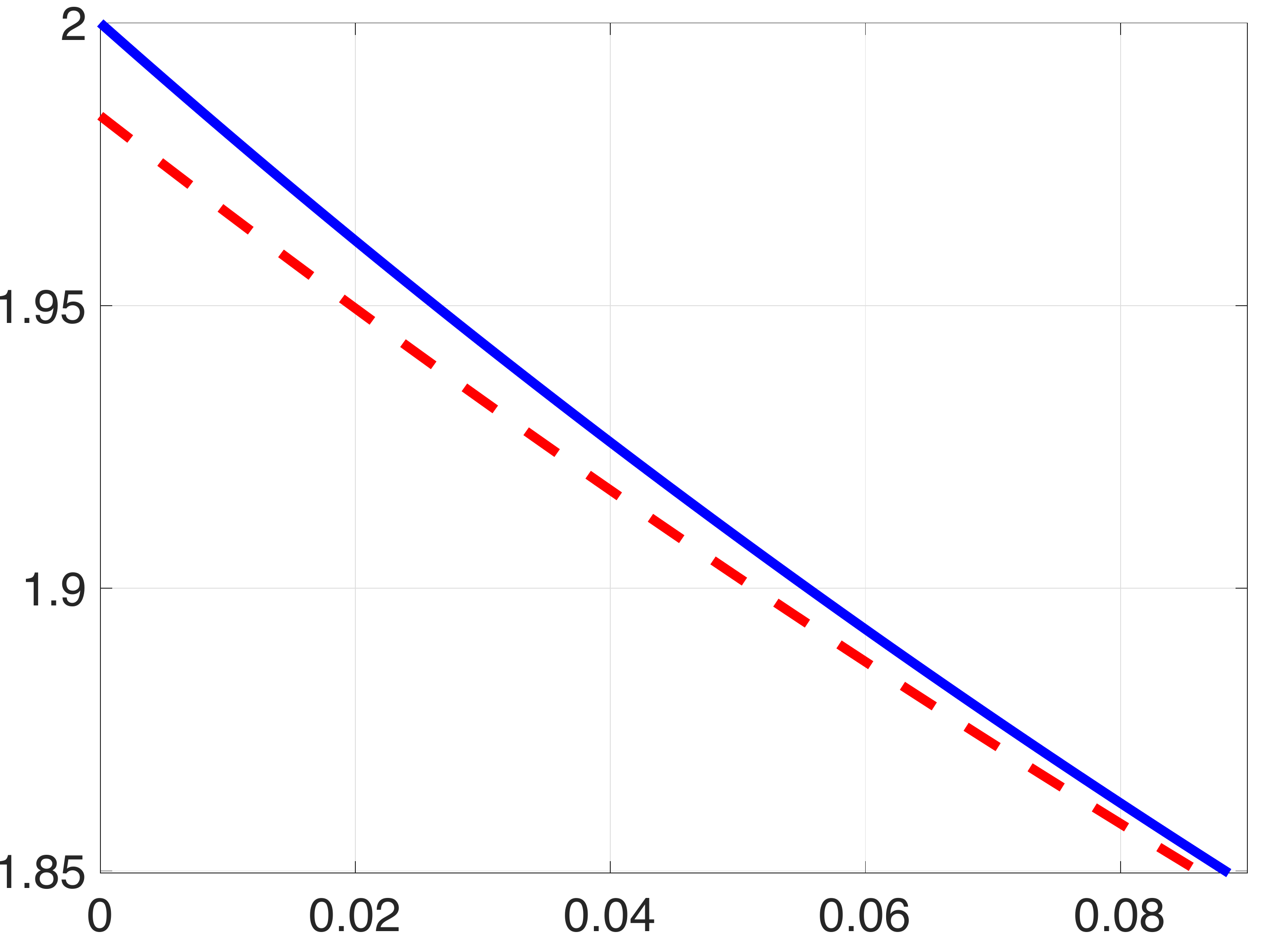} 
         \end{overpic}}
            \end{overpic}  \vspace{-2pt}
      \label{fig4}
      \begin{overpic}[scale=0.2]{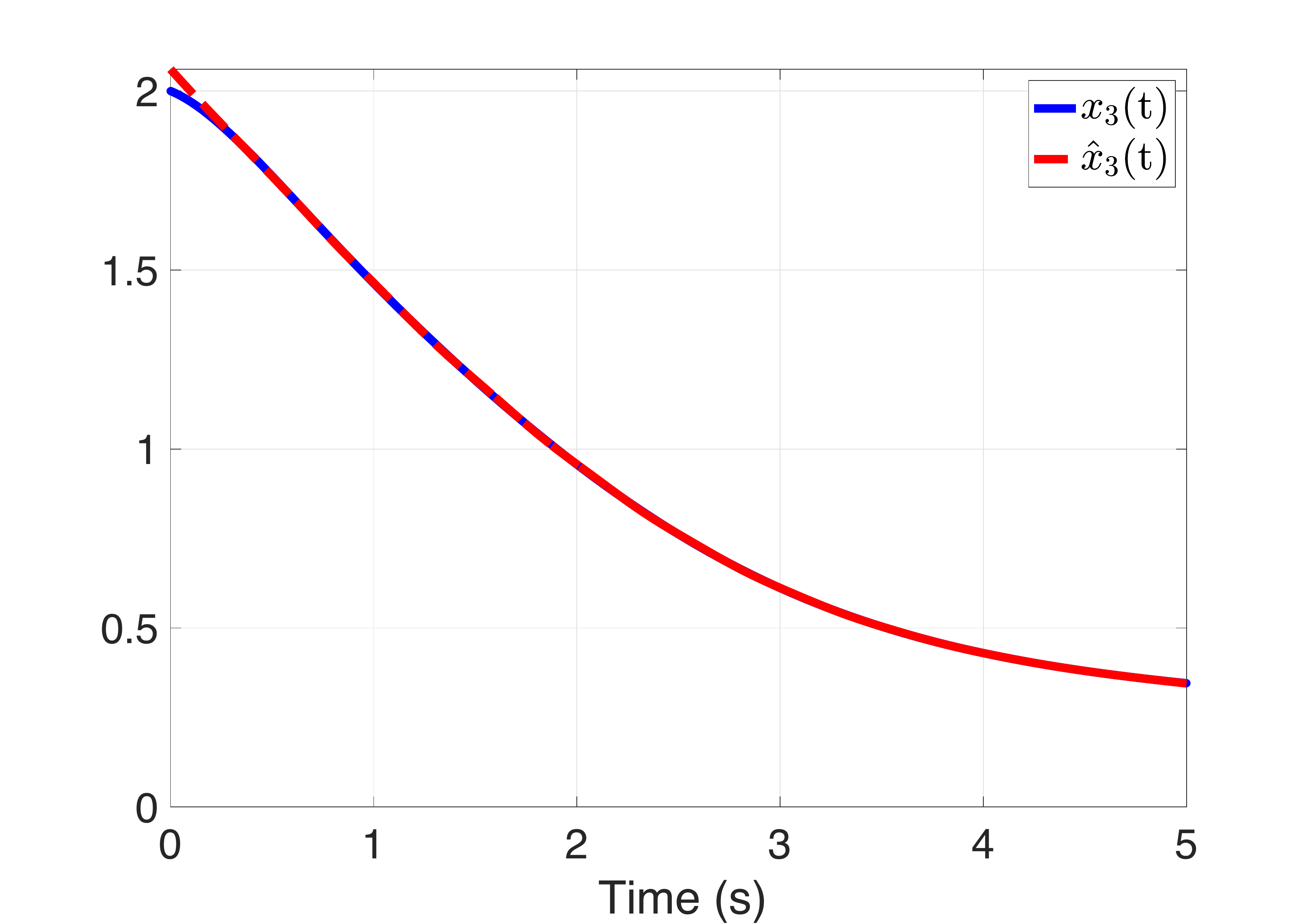}
      \put(40,45){\begin{overpic}[scale=0.08]{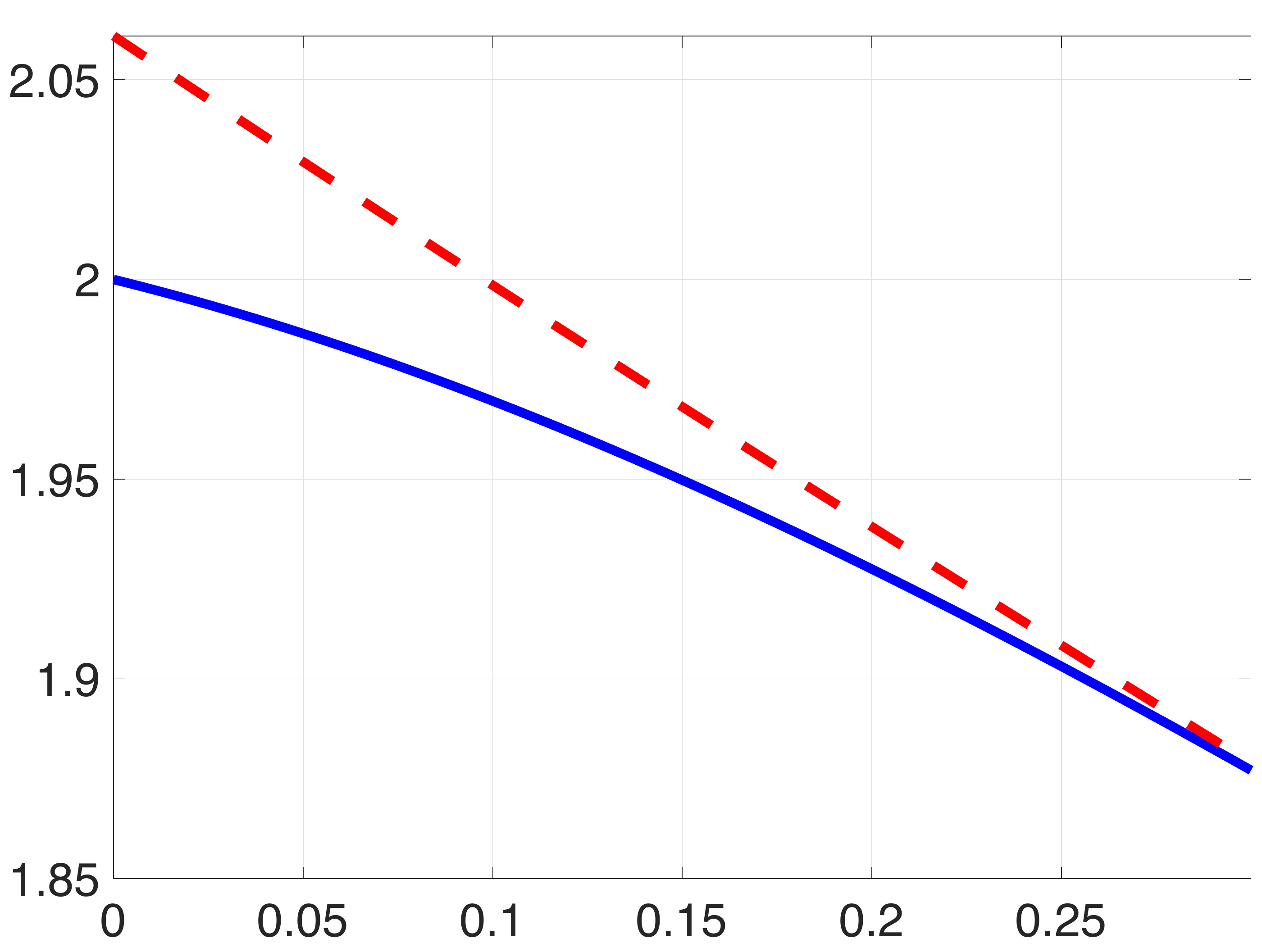} 
         \end{overpic}}
            \end{overpic}  \vspace{-2pt}
      \label{fig5}
     \begin{overpic}[scale=0.195]{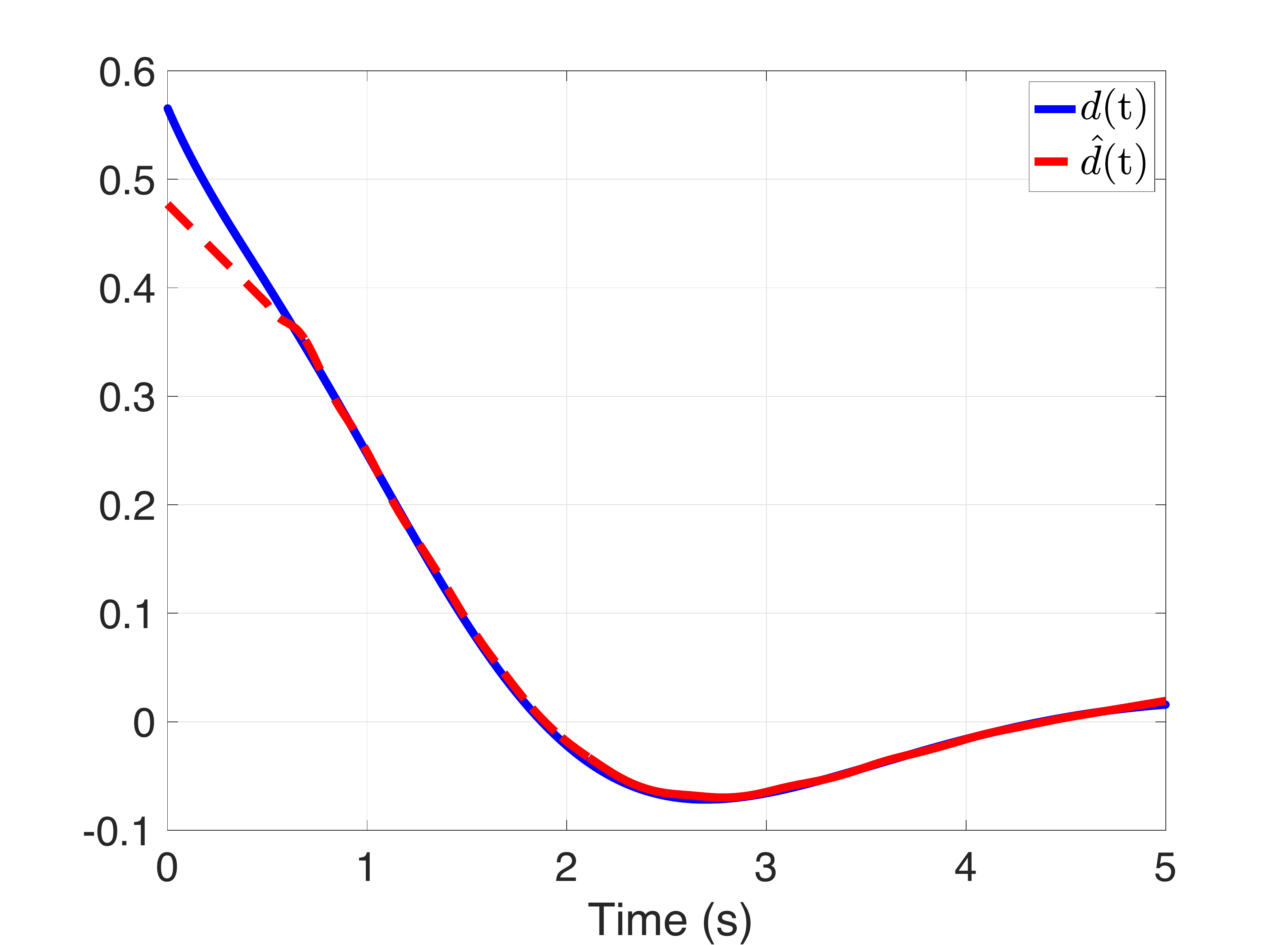}
     \put(45,-1.5){ \scriptsize Time (s)}
     \end{overpic}
      \caption{Online estimation of the states and the disturbance term. }\label{fig2}
\end{figure*}

\subsubsection{Online estimation}
The  normalized polynomial  modulating functions are adapted to the moving horizon scheme by considering
\begin{align}\label{eq57b}
  \bar{\phi}_i(\tau-t+h)=(t-\tau)^{(p_{x,k}+i)} (\tau-t+h)^{(p_{x,k}+S_k+1-i)}, \\ \notag
  \forall i=1,\cdots, S_k,  k=1,2,
\end{align}
\begin{equation}
\bar{\phi}_i(t)=(t-\tau)^{(p_d+i)} (\tau-t+h)^{(p_{x,k}+D+1-i)}, i=1,2, \cdots, D.
\label{eq58}
 \end{equation}  
 The moving integration window size was taken as $h=1$s, and the following parameters for the modulating function-based estimation were considered: $S_2=M_2=5$, $S_3=M_3=4$, $D=N=2$, $p_d=p_{x,1}=2$, $p_{x,2}=3$. \\
 Figure \ref{fig2} shows that modulating function-based estimator provides an accurate online estimation of both the states and the disturbance, which offers the potential of good closed-loop performance in case it is combined with a controller.

\subsection{Comparison with a second order sliding mode super twisting observer}
\noindent
We consider an unforced pendulum subject to Coulomb friction and external
disturbance given by the following second-order system \cite{Fridman2005}
\begin{equation*}
\left\{
\begin{aligned}
& \dot{x}_1=x_2 \\
& \dot{x}_2=-\frac{g}{L} \sin x_1-\frac{V_s}{J} x_2-\frac{P_s}{J} \operatorname{tanh}\left(\frac{x_2}{0.1}\right)+d(t)\\
& y=x_1
\end{aligned} \right.
\end{equation*}
The pendulum parameters have the following values $M=1.1, g=9.815, L=0.9, J=M L^2=$ $0.891, V_S=0.18, P_s=0.45$ and the disturbance term is given by 
\begin{equation*}
d(t)= 0.5 \sin(t) + 0.5 \cos(2t).
\end{equation*}
The state $x_2$ and the disturbance term $d$ are estimated online using the modulating functions in \eqref{eq57b} and \eqref{eq58}, and polynomial basis functions, where  $S_2=M_2=7$, $D=N=3$, $p_d=p_{x,1}=2$, and $h=1$s. The state estimation result is compared with the state estimation obtained by the second order sliding mode super twisting observer (STO) in \cite{Fridman2005} given by 
$$
\left\{ \begin{aligned}
& \dot{\hat{x}}_1=\hat{x}_2+1.5\left(f^{+}\right)^{1 / 2}\left|x_1-\hat{x}_1\right|^{1 / 2} \operatorname{sign}\left(x_1-\hat{x}_1\right) \\ 
& \dot{\hat{x}}_2=-\frac{g}{L_n} \sin x_1-\frac{V_{s_n}}{J_n} \hat{x}_2+1.1 f^{+} \operatorname{sign}\left(x_1-\hat{x}_1\right)
\end{aligned}\right.$$
where $f^{+}=6$ is the double maximal possible acceleration of the system\cite{Fridman2005}. The initial conditions are taken as $\hat{x}_1(0)=y(0)=x_1(0)$ and $\hat{x}_2(0)=0$.

\begin{figure*}[!t]
   \begin{minipage}[c]{0.45\linewidth}  
           \centering
      \begin{overpic}[scale=0.27]{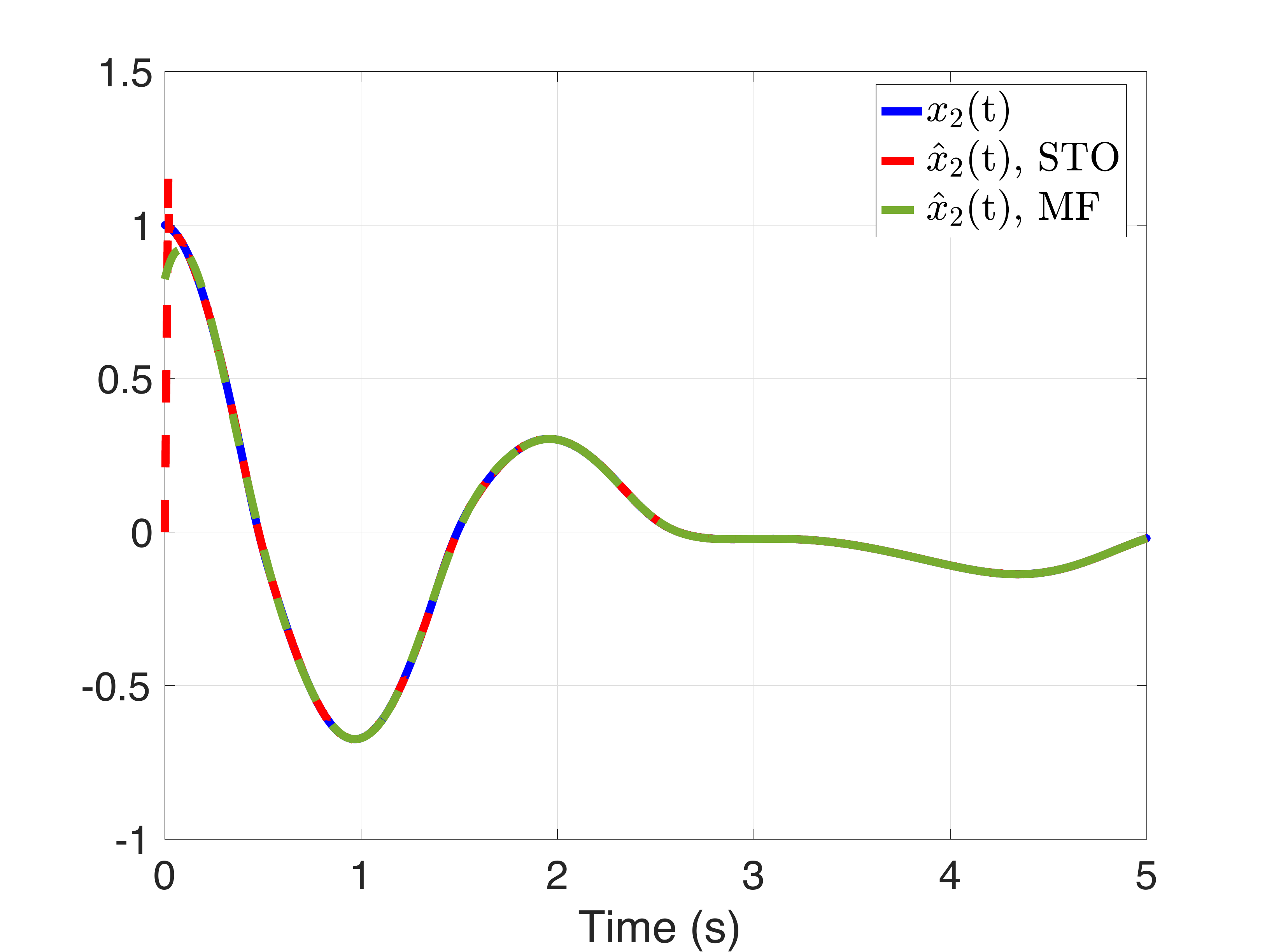}
      \put(20,0){\footnotesize  \textbf{a)}}
      \put(14,51){\begin{overpic}[scale=0.10]{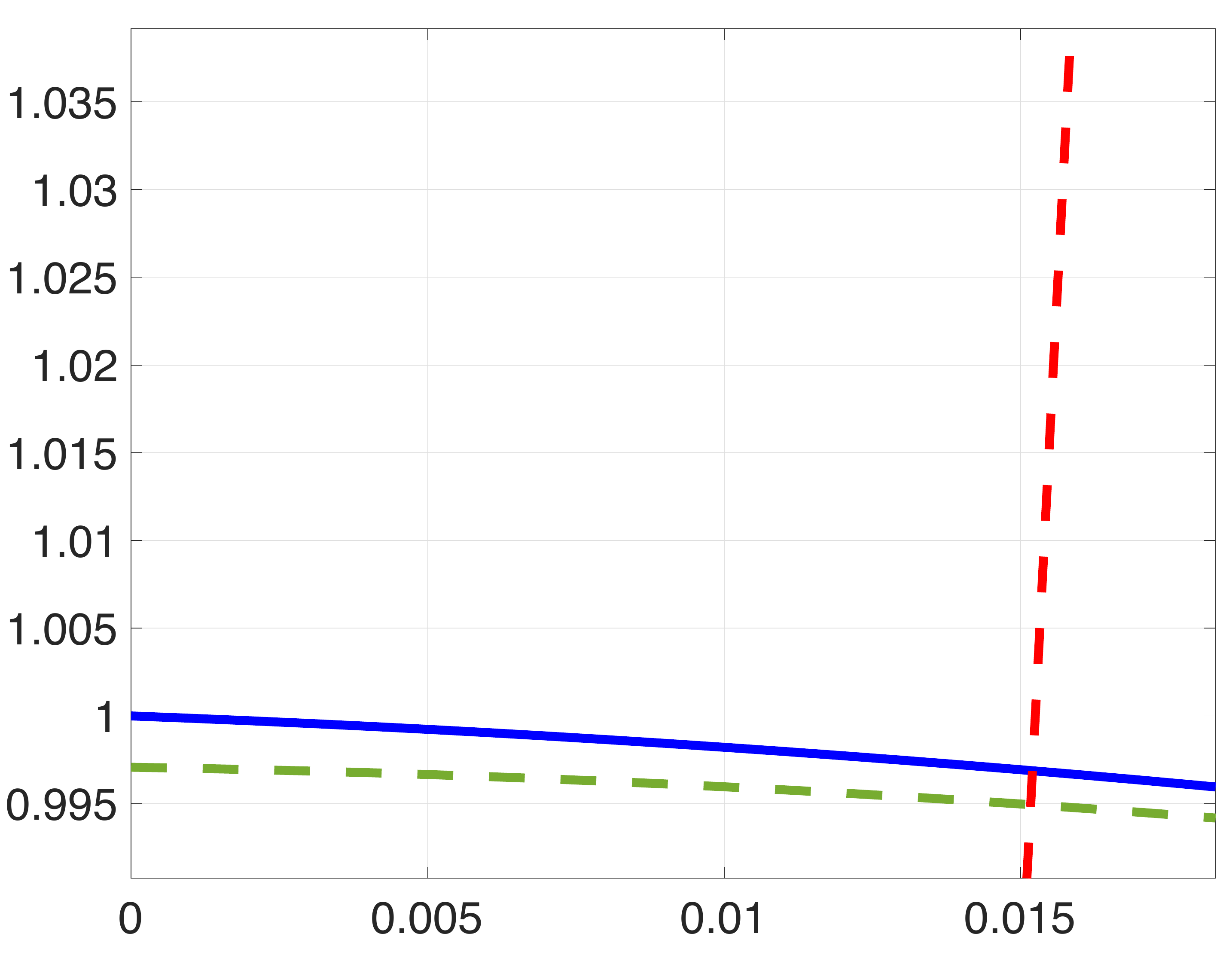} 
         \end{overpic}}
            \end{overpic}   
       \end{minipage}\hfill 
         \begin{minipage}[c]{0.48\linewidth}
         \centering
\begin{overpic}[scale=0.27]{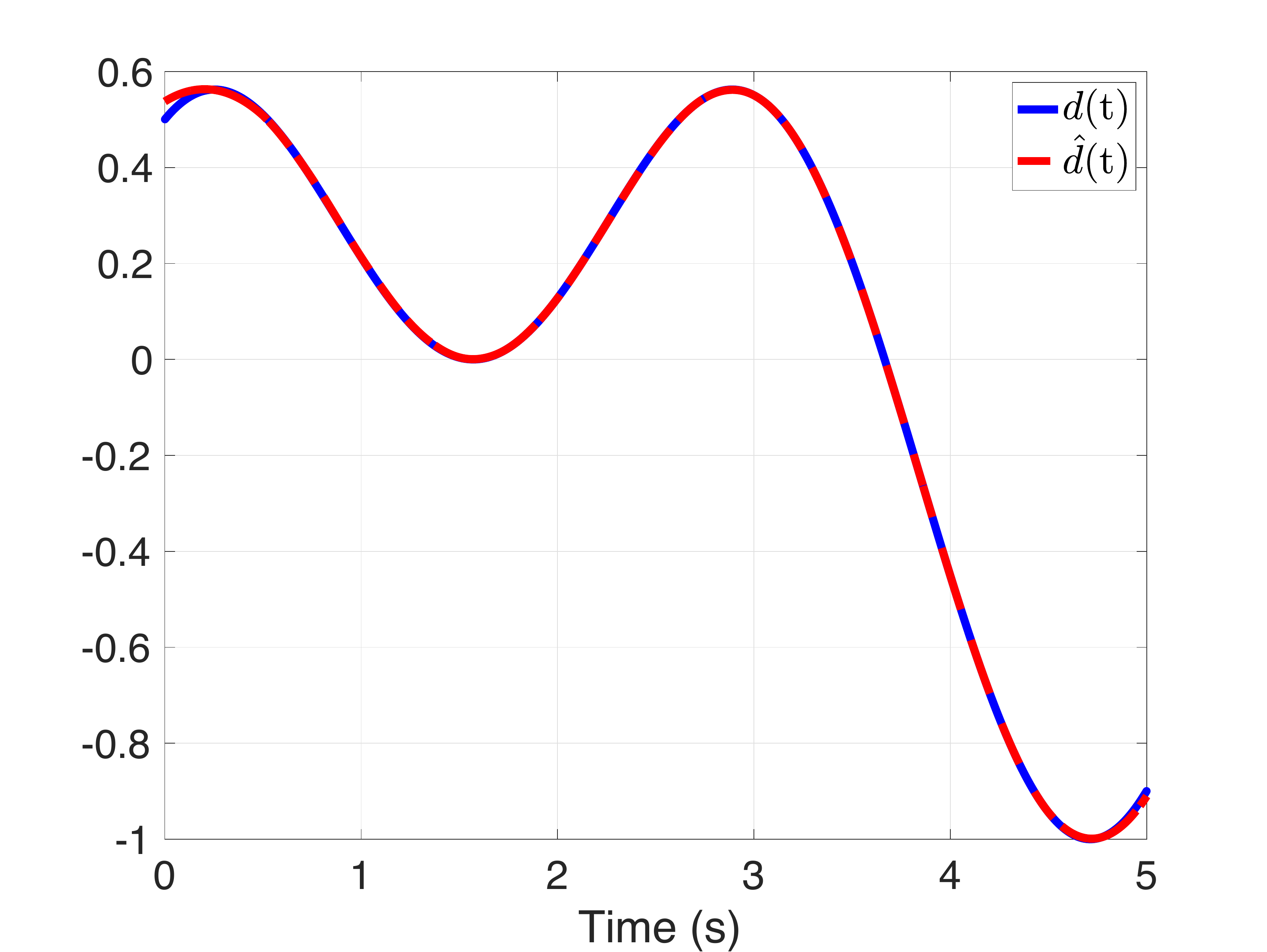}
\put(20,-1){ \footnotesize \textbf{b)}}
      \put(10,15){\begin{overpic}[scale=0.12]{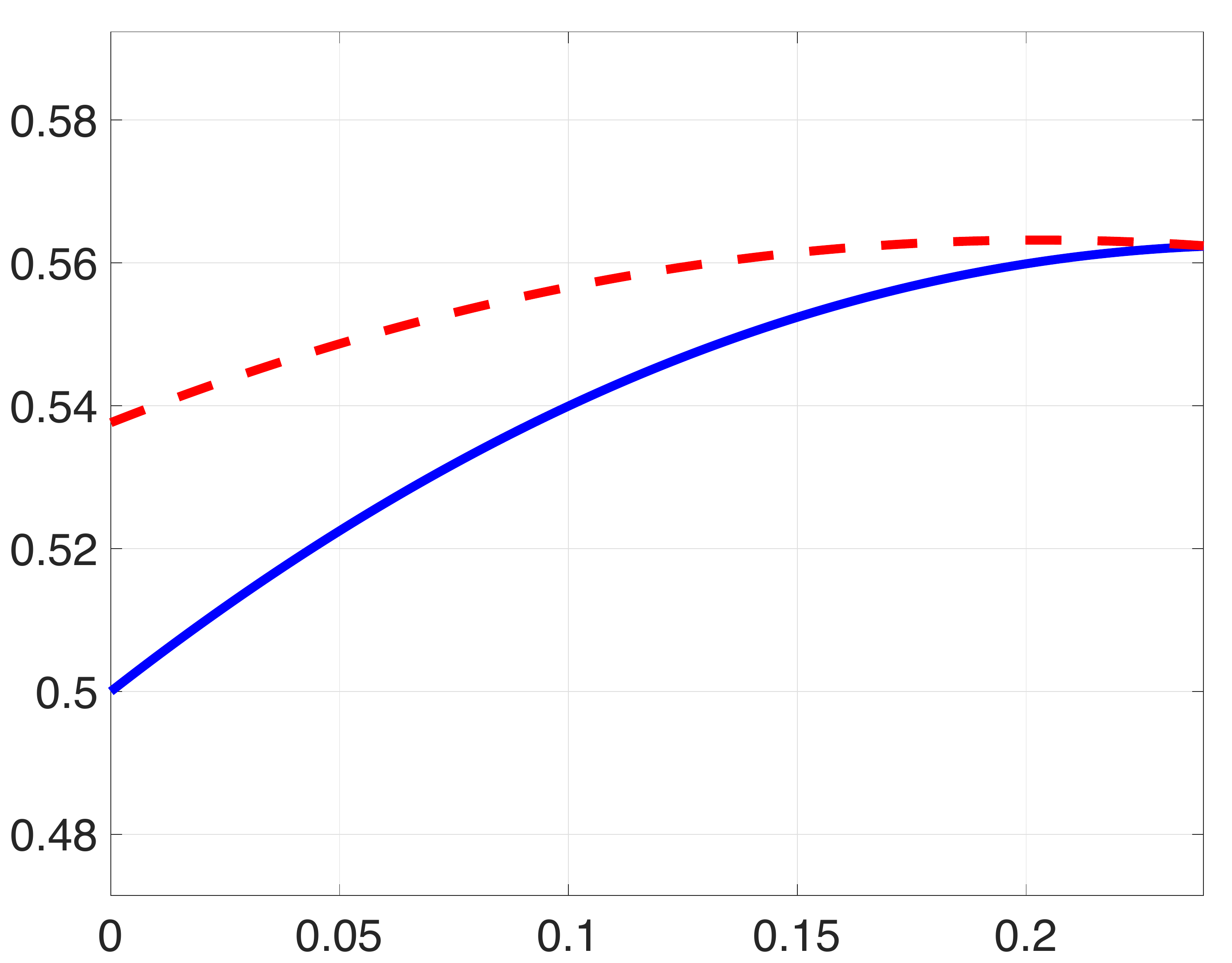} 
         \end{overpic}}
            \end{overpic}         
      \end{minipage} \vspace{0.2cm}          
      \caption{a) Online estimation of the state $x_2$ using MFBM and STO; b) online estimation of the disturbance $d$ using MF.} \label{fig7}
\end{figure*}

\noindent
The estimated state using Modulating Function Based Method (MFBM) and the Super Twisting observer (STO) are illustrated in Figure~\ref{fig7}\textbf{a)}. Both observers provide an accurate estimation of the state. Unlike the super twisting observer, the modulating function estimator does not require the initial condition which results in a faster convergence and a smaller relative estimation error as displayed in Table \ref{tab1}. In addition to the state estimation, the modulating function estimator allows estimating the disturbance $d$. In Figure~\ref{fig7}\textbf{b)}, one can see that the MFBM reconstructs accurately the disturbance.

\subsection{Robustness analysis}
\begin{figure*}[!t]
   \begin{minipage}[c]{0.45\linewidth}  
           \centering
      \begin{overpic}[scale=0.27]{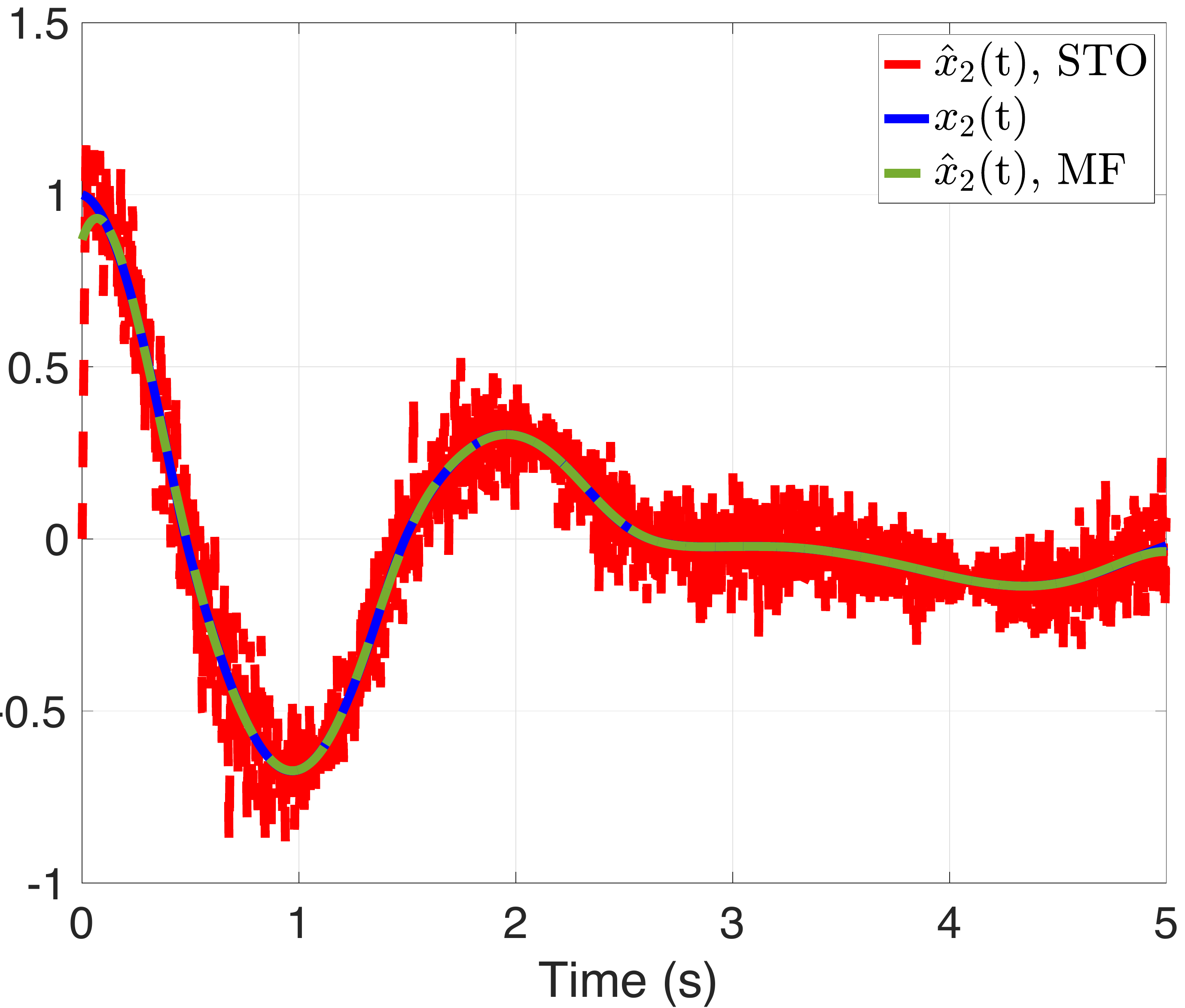}
      \put(20,0){\footnotesize  \textbf{a)}}
            \end{overpic}   
       \end{minipage}\hfill 
         \begin{minipage}[c]{0.48\linewidth}
         \centering
\begin{overpic}[scale=0.27]{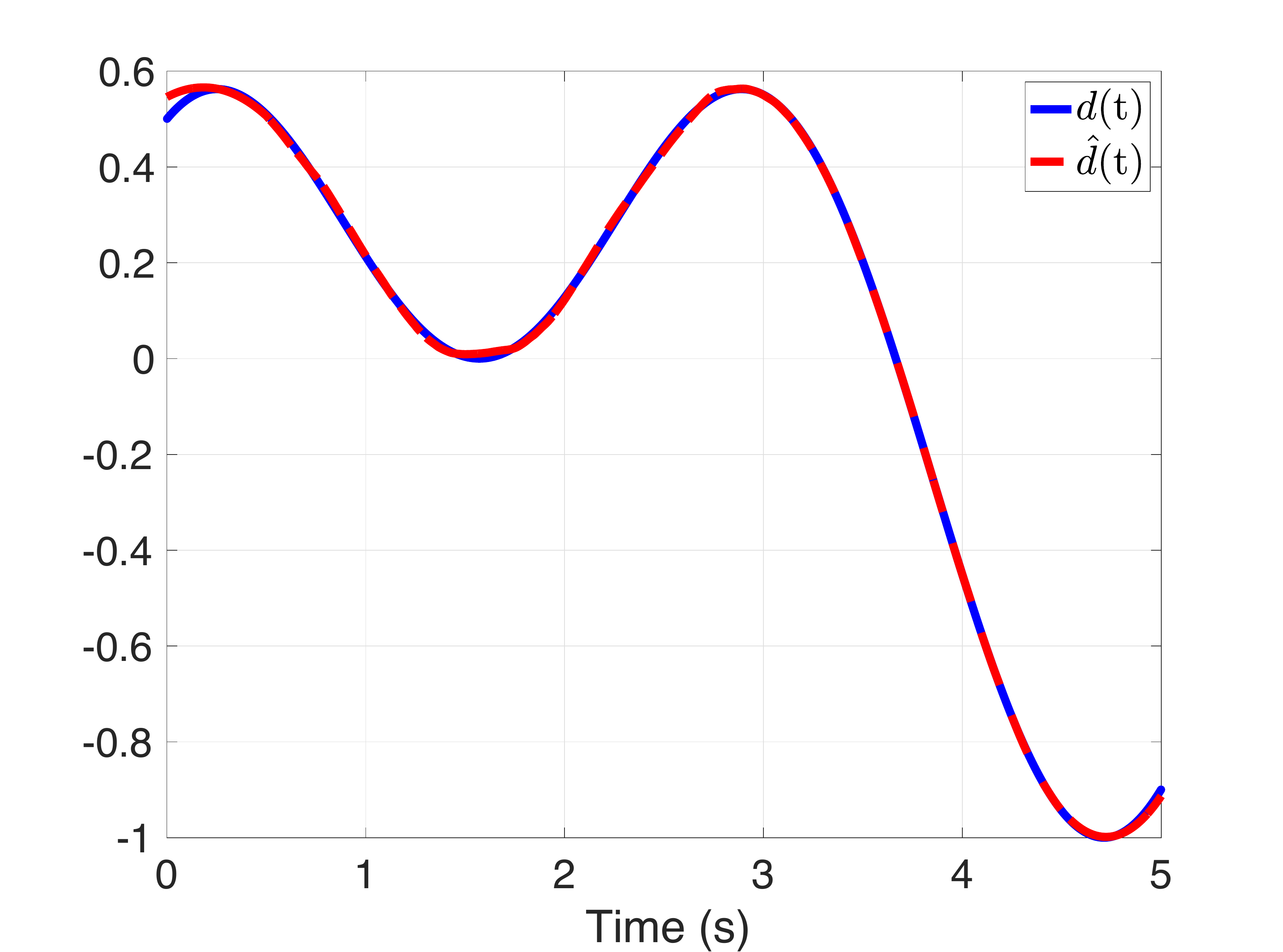}
\put(20,-1){ \footnotesize \textbf{b)}}
            \end{overpic}         
      \end{minipage} \vspace{0.2cm}          
      \caption{a) Online estimation of the state $x_2$ using MF and STO in presence of noise; b) online estimation of the disturbance $d$ in presence of noise.} \label{fig9}
\end{figure*}

\noindent
To evaluate the robustness of the proposed estimator against measurement noise, different levels of white Gaussian noise were added to the output $y$.  
Figure \ref{fig9}\textbf{a)} shows the state estimation with modulating function estimator and super twisting observer in presence of $1\%$ level of Gaussian noise. The state estimation with the MFBM is smoother and less affected by measurement noise. This is due  mainly to two factors: the result of Property \ref{property1} and the integral operator that reduce the effect of measurement noise. In Figure~\ref{fig9}\textbf{b)}, one can see that the MFBM reconstructs accurately the disturbance in presence of noise. Table~\ref{tab1} shows the state relative estimation error for different levels of noise. One can see the relative estimation error is smaller with the MFBM than with STO.

\begin{table}[ht]
\centering
\caption{Relative estimation error w.r.t different noise levels in \%}
\begin{tabular}[t]{|l|c|c|}
\hline
Noise level &  $\frac{|x_2-\hat{x}_2|}{|x_2|} $ with STO &  $\frac{|x_2-\hat{x}_2|}{|x_2|} $ with MF \\
\hline

0 $\%$ &10.43 & 1.41 \\

1 $\%$ & 28.13 & 1.81 \\

3 $\%$ & 33.29 & 2.07 \\

5 $\%$ & 45.58 & 2.92 \\

10 $\%$ & 49.67 & 7.55 \\
\hline
\end{tabular}
\label{tab1}
\end{table}

\section{Conclusion}
\label{conclusion}
\noindent
The present paper proposed a robust step-by-step non-asymptotic estimator based on modulating functions to estimate the states and disturbance term of nonlinear triangular systems. The robustness properties of the modulating function-based estimator are the result of the modulating operator and the main property of modulating functions that shifts the derivatives of the input-output towards the smooth modulating function which makes the proposed non-asymptotic estimator input-output derivative-free. Additionally, the modulating function-based method transforms the estimation problem from solving ordinary differential equations to solving a set of algebraic equations. Therefore the initial condition is not required which results in a faster convergence and a smaller estimation error.  The proposed observer was applied to an academic example of a third-order nonlinear triangular system to evaluate its performance where both offline and online estimation schemes were considered. Moreover, it was compared to the second-order super twisting sliding mode observer under different levels of measurement noise. While the sliding mode observer is more robust to model uncertainties, the modulating function estimator is more robust against measurement noise.      
 Future work will focus on extending the modulating function-based estimator to a more general class of nonlinear uncertain systems.

\bibliographystyle{unsrt}
\bibliography{ref.bib}

\end{document}